\documentclass[11pt]{article}

\usepackage{style}
\usepackage{tikz_style}

\usepackage[margin=1in]{geometry}
\usepackage{comment}

\newtheorem{theorem}{Theorem}
\newtheorem{lemma}{Lemma}

\title{Universal Dancing by Luminous Robots under Sequential Schedulers}
\date{}
\author{
    Caterina Feletti\footnote{Department of Computer Science, Università degli Studi di Milano, Italy, \textit{caterina.feletti@unimi.it}} 
    \and
    Paola Flocchini\footnote{School of Electrical Engineering and Computer Science, University of Ottawa, Canada, \textit{paola.flocchini@uottawa.ca}} 
    \and
    Debasish Pattanayak\footnote{Department of Computer Science and Engineering, Indian Institute of Technology Indore, India, \textit{debasish@iiti.ac.in}}
    \and
    Giuseppe Prencipe\footnote{Department of Computer Science, Università di Pisa, Italy, \textit{giuseppe.prencipe@unipi.it}} 
    \and
    Nicola Santoro\footnote{School of Computer Science, Carleton University, Canada, \textit{santoro@scs.carleton.ca}}
}

\begin{document}
    \maketitle

    \begin{abstract}
    The \probfont{Dancing} problem requires a swarm of $n$ autonomous mobile robots to form a sequence of patterns, aka perform a \emph{choreography}.
    Existing work has proven that some crucial restrictions on choreographies and initial configurations (e.g., on repetitions of patterns, periodicity, symmetries, contractions/expansions) must hold so that the \probfont{Dancing} problem can be solved under certain robot models.
    Here, we prove that these necessary constraints can be dropped by considering the $\LUMI$ model (i.e., where robots are endowed with a light whose color can be chosen from a constant-size palette) under the quite unexplored \emph{sequential scheduler}.
    We formalize the class of \probfont{Universal Dancing} problems which require a swarm of $n$ robots starting from any initial configuration to perform a (periodic or finite) sequence of arbitrary patterns, only provided that each pattern consists of $n$ vertices (including multiplicities).
    However, we prove that, to be solvable under $\LUMI$, the length of the feasible choreographies is bounded by the compositions of $n$ into the number of colors available to the robots.
    We provide an algorithm solving the \probfont{Universal Dancing} problem by exploiting the peculiar capability of sequential robots to implement a distributed counter mechanism.
    Even assuming non-rigid movements, our algorithm ensures spatial homogeneity of the performed choreography.
\end{abstract}

    \textbf{Keywords: }{Luminous Robots, Sequence of Patterns, Pattern Formation, Sequential Scheduler.}

\section{Introduction}

\subsection{Framework and Background}

In the 2020 Olympic Opening Ceremony, a swarm of thousands of drones combined to display an artistic performance in the sky of Tokyo.
While it is true that such a technological spectacle has required an engineering effort, it is evident that it could not have been achieved without an accurate algorithmic study.
In that case, the drones had to combine in a sequence of patterns (e.g., the Olympic rings), thus performing a sort of choreography, ``dancing'' in perfect synchrony, and avoiding collisions.

In the last two decades, scenarios like this have been formally modeled and studied to design algorithmic techniques intended for the emerging field of \emph{swarm robotics}.
Besides entertainment and performance, swarm robotics can be adopted in multiple domains (e.g., agricultural, aerospace) and for different promising purposes (e.g., exploration, data collection, security, and rescue).
Typically, a swarm is modeled as a decentralized and distributed system where multiple computational entities (aka robots) have to arrange on the environment to solve a common task collaboratively.

The \emph{Look-Compute-Move} (LCM)  is one of the main models used to describe these systems \cite{FloccPS19}.
Generally, a swarm is modeled as a set of punctiform mobile robots that can act on the Euclidean plane.
The robots are idle by default, and 
are repeatedly activated by a scheduler.
As soon as activated, a robot performs a  LCM cycle: it takes the snapshot of the system (\emph{Look}), it executes a deterministic algorithm calculating a position (\emph{Compute}), and eventually it moves straight towards the computed position.
By repeating the LCM cycle infinitely, the robots perform a given task (aka solve a problem): typically, robots are required to arrange in the environment to satisfy some conditions (e.g., \probfont{Scattering} \cite{DieudP09,IzumiKBT18}) or according to some patterns (e.g., \probfont{Pattern Formation} \cite{BramaT16,CicerDN19a,CicerSN21,DPV10,FloccPSW99,FloccPSW08,Prencipe19,YamauY14,VaidyST22}, \probfont{Gathering} \cite{BramaT15,ChaudM15,CieliFPS12,KameiLOT11,Flocc19_Gathering,PattaAM21,PattaMHM19}).
To design algorithms as robust as possible---namely, which can be executed in 
very limited systems---
the literature considers robots that are: \emph{anonymous} (no internal ids), \emph{indistinguishable} (no external ids), \emph{autonomous} (no central control), and \emph{homogeneous} (they execute the same algorithm).
Typically, robots are assumed to be partially or totally disoriented: there might not be agreement between their local coordinate systems, on 
the unit of distance, or on chirality (i.e., clockwise orientation of the plane).
Other limitations may concern robots' movements, which can be \emph{rigid} or \emph{non-rigid} according to the impossibility or possibility of being stopped by an adversary, respectively.

In the early literature, 
starting from 
the pioneering work of \cite{SuzukY99}, robots are assumed to be \emph{oblivious} and \emph{silent}, i.e., with neither a persistent memory nor communication means.
This very restricted LCM sub-model, called $\OBLOT$, has been widely studied to both explore its power and delimit its limitations regarding computability.
Not surprisingly, not all problems can be solved without memory or communication.
The more recent sub-model $\LUMI$ has been introduced by Das et al.~\cite{DasFPSY16} to overcome such limitations: here, robots are assumed to be \emph{luminous}, i.e., endowed with a persistent register, called {\em light}, whose value, called {\em color}, can be updated at each LCM cycle choosing from a $O(1)$-size palette of colors.
Being visible to all robots in the swarm, the robot's light serves as an internal persistent memory and an external communication means.

Nevertheless, it is not just the presence/absence of lights that changes the computing power of a swarm.
It has been proven that the setting under which robots are activated and synchronized can affect their power.
In the \emph{synchronous} settings, time is divided into atomic  discrete \emph{rounds}; at each round  an arbitrary non-empty subset of the robots is activated and they execute  their LCM cycle in perfect synchrony. 
The most  general synchronous setting is called \emph{semi-synchronous} ($\ssynch$) and it has been widely studied together with its special case \emph{fully synchronous} ($\fsynch$), where all the robots are activated simultaneously at every round.
In the even more general \emph{asynchronous} setting ($\asynch$), the robots are activated without any assumptions on 
the duration of a LCM cycle (except that is finite) nor on the synchronization among them.
Note that under $\ssynch$ (and thus $\asynch$), robots are unaware of the activation scheduling of the swarm, which is assumed to be made by an adversarial scheduler that, however, operates under the \emph{fairness condition}, stating that each robot is activated infinitely often within finite time.

For years, the literature has mainly focused on these three settings; notably, it has studied the computational power and limits of a model $X^{S}$, where $X\in \{\OBLOT, \LUMI\}$ represents the memory/communication setting while $S\in \{\fsynch, \ssynch,\asynch\}$ represents the synchronization setting \cite{BuchFKPSW21,DasFPSY16,FelettiMMP25,FloccSSW23}.

On the sidelines, the \emph{sequential} setting (\seq), which activates only one robot at a round, remained nearly unexplored until a few years ago.
Recently, some works have addressed this setting and its sub-settings (e.g., the famous \emph{round-robin}), especially to investigate fault-prone swarms \cite{ClementeF25,DefagoGMP06,DefagPT19}, and  the well-known \probfont{Pattern Formation} and \probfont{Gathering} problems \cite{FNPPS25,FreiW24}.

\subsection{Pattern Formation}
\probfont{Pattern Formation}  refers to the abstract problem of the robots rearranging themselves so that, within finite time, the set of their locations in the plane (the {\em configuration}) forms (irrespective to rotation, reflection, translation or scaling) a given geometric figure (the {\em pattern}), and the robots no longer move.


Which patterns are formable depends on many factors, first and foremost on the model $X^{S}$,  where $X\in \{\OBLOT, \LUMI\}$ and $S$ is the adversarial scheduler.
For a given model $X^{S}$, the main focus has been on determining which patterns are formable from a given initial configuration, and, more importantly, which patterns are formable from every initial configuration  (aka {\tt Arbitrary Pattern Formation}). 
All these problems and questions have been extensively studied in the LCM model under several different assumptions (e.g., see \cite{BramaT16,CicerDN19a,CicerSN21,DPV10,FloccPSW99,FloccPSW08,Prencipe19,VaidyST22,YamauY14}).

Clearly, to form a pattern, there must be at least as many robots as there are points in the pattern; we shall call this the {\em trivial assumption} and assume to hold.
Most of the algorithmic research has however relied on a variety of non-trivial assumptions, such as: restricting the type of allowed initial configurations (e.g., the robots occupy distinct initial locations); imposing restrictions on the number of robots or the size of the pattern (e.g., the size of the swarm is a prime number, or it is equal to the number of points in the pattern); requiring specific symmetry relationship between the initial configuration and the pattern; the robots have some agreement on coordinate systems or on chirality; the movements of the robots are rigid; one of the robots is visibly different from all the others (i.e., there is a {\em leader}), etc. (see, e.g.,~\cite{CicerDN19a,CicerDN19,CicerSN21,FloccPSV17,FloccPSW08,SuzukY99,YamasS10}).

The most general problem in this class, \probfont{Universal Pattern Formation} (\probfont{UPF}), requires the robots to form {\em every} arbitrary pattern given in input,  starting from {\em any} arbitrary initial configuration,   regardless of the number of robots, and of the number of points in the pattern, under just
the trivial assumption. As for the power of the robots, the only requirement is strong multiplicity detection (i.e., robots can detect the
exact number of robots occupying the same location), which is required by the nature of the problem.
This problem is generally unsolvable in $\OBLOT$ under $\fsynch$ (and, thus, $\ssynch$) even with a leader, full agreement on the coordinate systems, and rigid movements.
However, it has recently been shown that, except for point formation, {\tt Universal Pattern Formation} is solvable in $\OBLOT$ under the \emph{sequential} setting without any additional assumption \cite{FNPPS25}.  
This implies that {\tt UPF}, including point formation, is solvable in $\LUMI$ under the \emph{sequential} schedulers, i.e., in $\LUMI^\seq$. 
This result brings to light the strong computational power that robots have under the sequential  schedulers in regards to
pattern formation problems.  

Interestingly,  \probfont{Pattern Formation} can be a primitive task to be repeatedly achieved in a more complex task.
Examples of such more complex tasks are \probfont{Flocking}, requiring the robots to construct a pattern and then maintaining it while moving  \cite{CanepDIP16,GervasiP04}, and \probfont{Dancing} ---as nicely named in \cite{DasFPS14}--- requiring the swarm to form a given ordered sequence of patterns (or {\em choreography}) \cite{DasFPS20,DasFSY15}, which is the subject of our investigation.

\subsection{Sequence of Patterns: Dancing}

In the study of pattern formation by swarms of robots, an immediate question naturally arose: without persistent memory and direct communication (as in $\OBLOT$), can a swarm form more than one pattern in a specified order?
In other words, can the \probfont{Dancing} problem be solved under $\OBLOT$?

This question was surprisingly  answered in the positive in \cite{DasFSY15}, where it was shown that an $\OBLOT^\ssynch$ swarm can 
indeed form  sequences of patterns. The possibility to ``perform a choreography'' $\choreo$ is  however subject 
to several constraints: on the patterns (i.e., they  must have the same number of vertices and   degree of symmetry\footnote{Conditions on the symmetry of the patterns and of the initial configuration derive from the fact that symmetric robots can be activated simultaneously under $\ssynch$ (and, thus, $\asynch$) and they may perform a symmetric movement, thus never “breaking” the original symmetry.}  
and must be all distinct); on the number of robots (i.e., it must be equal to the number of vertices of the patterns); on the
 initial configuration (i.e., the robots must start from distinct locations and their positions must have same symmetricity as the patterns); and on the structure of the choreography (i.e., the sequence must  be periodic).
Under these inevitable constraints, an algorithm was presented  that allows oblivious robots to perform a choreography employing a peculiar technique,  based on the related distances of selected robots, to collectively memorize the index of the next pattern to be formed  \cite{DasFSY15}; this algorithm assumes chirality agreement  and rigid movements.


The impact that the limited ability of memory and communication provided by $\LUMI$ has on the feasibility of \probfont{Dancing} was studied next \cite{DasFPS20}. 
It was shown that, in $\LUMI$, the feasible sequences are subject to fewer constraints; more precisely, a swarm can perform a periodic choreography containing pattern repetitions and contractions/expansions, starting from any configuration with all robots at distinct locations, subject to a weaker conditions of symmetry on the patterns and initial configuration.
The authors show how swarms in $\LUMI$ can perform any  choreography feasible  in $\LUMI$  even under the $\asynch$ scheduler, and  with non-rigid  movements
\cite{DasFPS20}.

These studies on \probfont{Dancing} have brought to light an important fact in the passage from \probfont{Pattern Formation} to \probfont{Dancing}.
Let ${\cal P}(X^S)$ be the set of patterns formable in model $X^{S}$.
The results of \cite{DasFPS20,DasFSY15} imply that, given a set $P\subset {\cal P}(X^S)$ of feasible patterns, it does not mean that any periodic sequence of patterns chosen from $P$ is a feasible choreography under $X^S$.
In other words, {\em a  sequence of feasible patterns is not necessarily a feasible sequence}.

In this paper, we extend the current investigations  on  \probfont{Dancing}  and on  the computational power of swarms
under sequential schedulers, 
 and establish feasibility and complexity results on \probfont{Dancing} in  $\LUMI^\seq$.

\subsection{Contributions: Universal Dancing}

We investigate if and under which constraints \probfont{Dancing} is solvable under the luminous-sequential model, establishing several results.

As mentioned previously, \probfont{Universal Pattern Formation} is solvable in $\LUMI$ under the {sequential} setting without any additional assumption \cite{FNPPS25}; that is the class of patterns formable in $\LUMI^\seq$ includes every possible pattern.

In this paper, we prove that, in  $\LUMI^\seq$, {\em every sequence of patterns is formable}, independently of any property (e.g.,  symmetry, size, shape, etc.)  of the patterns or of the initial configuration, whether the sequence is periodic or not.
In other words, we prove the sequential scheduler enables a swarm of luminous robots to solve what we shall call the \probfont{Universal Dancing} problem.

Our proof is constructive: we develop a solution algorithm, analyze its properties, and establish its correctness.
The main ingredient of our solution is the implementation, by the robots themselves,  of 
 a distributed colored counter, which allows them to recognize at all times the current pattern being formed  
 in the choreography.     This new technique exploits the sequential nature of the scheduler
   with a careful use of colors  and of an extended composition Gray code.  
  The   formation of each  individual pattern is  based on a novel technique which makes the robots match with their respective target points exploiting properties of Dyck words on the Boolean alphabet.
Moreover, unlike previous solutions, our algorithm allows a swarm to perform choreographies of finite length. 
Our solution tolerates non-rigid movements.

Our results clearly show the strong computational power that robots have under the sequential schedulers in regard to pattern formation problems, confirming and extending to the $\LUMI$ model the evidence found for $\OBLOT$ in \cite{FNPPS25}.

It is noteworthy that, besides the removal of the classical constraints, our solution guarantees an interesting spatial property which provides homogeneity to the whole choreography performance: the smallest circles enclosing the patterns of the choreography (except for patterns with only two or three points) formed by the swarm are concentric. 

    \section{Preliminaries}\label{sec:preliminaries}
\subsection{Robots}
We consider a swarm of $n$ mobile computational entities, \emph{aka} robots, $\swarm = \{r_1,\dots, r_{n}\}$, acting in the Euclidean plane $\Reals^2$. 
Robots are assumed to be autonomous (i.e., without any central control), anonymous and indistinguishable (i.e., devoid of internal and external ids), homogeneous (i.e., they execute the same algorithm), and punctiform.
Each robot $r$ has its own local coordinate system $\COO_r$.
We assume that robots are \emph{disoriented}, thus their local coordinate systems may be totally different (no global agreement on axes, unit distance, origin, chirality); moreover, the local coordinate system may change from one robot's activation to the next one.
For simplicity, we assume each $\COO_r$ is egocentric, i.e., it defines its origin in the current position of $r$.
We consider the $\LUMI$ model (i.e., \emph{luminous}), where each robot $r$ is embedded with a persistent light $\light_r$ whose color can be chosen from a constant-size palette $\palette$.
We assume that $\palette$ always contains the default color $\off$, which all robot lights are initialized to.
Moreover, $r$ is provided with a sensing system that allows it to perceive the positions (according to its current $\COO_r$) and colors of all the robots in the swarm.
By default, a robot is idle. 
A scheduler is responsible for the robots' activation times.
Once activated, a robot $r$ performs a \emph{Look-Compute-Move} cycle.
\begin{itemize}
    \item \emph{Look:} $r$ obtains a snapshot $\sigma$ of the positions according to $\COO_r$ occupied by all the robots, and the corresponding lights' colors.
    \item \emph{Compute:} $r$ executes the shared algorithm, say $\algo$, giving it the snapshot $\sigma$ in input, and computes a destination as a point on the Euclidean plane and a new color for its light $\light_r$.
    Formally, it computes $\algo(\sigma) = (p',c)$, with $p'\in\Reals^2$ and $c\in\palette$.
    Then, $r$ updates the color of $\light_r$ to $c$.
    \item \emph{Move:} $r$ moves to the computed destination $p'$ traveling along a straight trajectory. If the computed destination is its current position, then it stays still (\emph{null movement}).
\end{itemize}
We assume \emph{non-rigid} movements: each robot may be stopped by an adversary before reaching its destination, but always after having traveled at least a distance $\delta$, where $\delta>0$ is arbitrarily small.
The guaranteed minimum distance $\delta$ is a fixed constant, but unknown to the robots.
\subsection{Scheduler}
We consider the sequential class of schedulers, abbreviated as $\seq$.
Time is logically divided into discrete and atomic rounds.
In each round, exactly one robot is activated, and it completes one Look-Compute-Move cycle in that round. 
The sequence of activation of robots is adversarial, in the sense that, an adversary controls the scheduler. 

Formally, a {\seq} scheduler $\AS$ defined for a swarm $\swarm$ is a function $\AS:\Nat \to \swarm$ which defines for any instant time $t\in\Nat$ the activated robot. 
When needed, we can replace the function-like notation of $\AS$ and define it through the \emph{activation sequence} $\AS(0)\AS(1)\dots$.
The scheduler satisfies the \emph{fairness} constraint, i.e., for any $r\in\swarm$ and any $t\in \Nat$, there is a time $t'>t$ such that $\AS(t') = r$.
Time is measured in \emph{epochs}: the first epoch starts with the first activation of $\swarm$;  each epoch ends as soon as all robots have been activated; the next epoch starts with the next activation. 
\subsection{(Sequences of) Patterns}
\subparagraph*{Patterns and vertices.}
Let us consider an absolute\footnote{Unknown by the robots.} coordinate system ${\GCOO}$ on the Euclidean plane $\Reals^2$; unless differently specified, we consider the positions of the points of the plane according to ${\GCOO}$. 
We indicate with $\dist{\cdot,\cdot}$ the Euclidean distance between two points.

We define a \emph{pattern} ${\Pi} = \multiset{v_1,\dots, v_n}$ as a multiset\footnote{We will use the notations $\multiset{}$ and $\{\}$ to indicate a multiset and a set, respectively.} of positions with $v_i\in\Reals^2$.
We say that each $v_i$ is a \emph{vertex} of the pattern.
We define the \emph{shape} of $\Pi$ as the set $shape(\Pi)$ containing the unique positions in $\Pi$.

A pattern $\Pi$ can belong to one of the four classes: $\Pi\in\Point$ if $shape(\Pi)$ is a singleton; $\Pi\in\TwoPoints$ if $|shape(\Pi)| = 2$; $\Pi\in\ThreePoints$ if $|shape(\Pi)| = 3$; otherwise, $\Pi\in\NPoints$.
We denote with $\NLine$ the special subclass of $\NPoints$ where the positions of $\Pi$ are aligned.

We denote with $\Pats{{\Pi}}$ the set of patterns similar to ${\Pi}$, i.e., all the patterns obtained from ${\Pi}$ by any composition of non-degenerate similarity transformations: uniform scaling with a non-zero scale factor, translation, rotation, and reflection.
Formally, $\Pi' \in \Pats{\Pi}$ if there exists a non-degenerate similarity transformation $\tau:\Pi\to \Pi'$.

It will be convenient to define a pattern $\Pi$ using the set notation, so that $\Pi=\{(v_1,\card_1), \dots, (v_m,\card_m)\}$, where $\card_i\in\Nat_{>0}$ indicates the multiplicity of the relative vertex and where $\sum_{j=1}^{m}\card_j=n$.
We refer to $(v_i,\card_i)$ as a \emph{multivertex} of $\Pi$.
\subparagraph*{Configurations and SEC.}
Let us consider a swarm of robots $\swarm=\{r_1,\dots,r_n\}$.
A \emph{configuration} of $\swarm$ at time $t$ is the tuple $C(t)=((p_1,c_1),\dots,(p_n,c_n))$ containing, for each robot $r_i\in\swarm$, its position $p_i\in\Reals^2$ and its color $c_i$ at time $t$.
In the \emph{initial configuration} $C(0)$, all the colors are set to $\off$; no other assumptions are made.
We will omit the time $t$ when no ambiguity arises.
Given a configuration $C$, we denote with $C_{|\Reals^2}$ the multiset containing only the positions in $C$.
We say that a multivertex $(v_i,\card_i)$ is \emph{saturated} if exactly $\card_i$ robots lie on $v_i$ in $C$; on the contrary, we say that it is \emph{unsaturated} (\emph{oversaturated}, resp.) if it is covered by fewer than (more than, resp.) $\card_i$ robots.
We say that $\swarm$ forms a pattern $\Pi$ at time $t$ if $C_{|\Reals^2}\in\Pats{\Pi}$.

Given a pattern $\Pi$, we indicate with $SEC(\Pi)$ the \emph{smallest enclosing circle} of $\Pi$ and with $\centersec{\Pi}$ the center of $SEC(\Pi)$.
We remind that the SEC of a set of points is unique, and at least three (or two antipodal) points of the set lie on its perimeter.
With a slight abuse of notation, we use $SEC(\cdot)$ and $\centersec{\cdot}$ also for configurations: thus, given a configuration $C$, we indicate with $SEC(C)$ and $\centersec{C}$ the smallest circle enclosing the points in $C_{|\Reals^2}$ and its center, respectively.
%
\subparagraph*{Choreographies.}
We define a \emph{sequence of patterns}, also known as \emph{choreography}, as $\choreo=(\Pi_0,\dots,\Pi_{q-1})^x$ where each $\Pi_i$ is a pattern and $x\in\{1,\infty\}$.
It must be $\Pi_{i+1}\notin \Pats{\Pi_i}$; if $\choreo$ is \emph{periodic} (i.e., if $x=\infty$), then it must also hold that $\Pi_0\notin\Pats{\Pi_{q-1}}$.
If $\choreo$ is periodic, we assume that the sequence $\Pi_0,\dots,\Pi_{q-1}$ cannot be written as $(\Pi_0,\dots,\Pi_{h-1})^{\frac{q}{h}}$ for any $h<q$.
We say that $q$ is the \emph{length} (\emph{period}, resp.) of $\choreo$ if $x=1$ ($x=\infty$, resp.).


We say that $\swarm$ \emph{performs} $\choreo$ if, for any $j=1,\dots,x$, there exists a series of increasing finite times $\{t_{i+(j-1)q}\}_{i=0,\dots,q-1,j=1,\dots, x}$ such that the swarm forms $\Pi_i$ at time $t_{i+(j-1)q}$.
If $x=1$, then the swarm must remain still after time $t_{q-1}$ (i.e., after having formed the last pattern of the choreography).
Obviously, a choreography $\choreo$ must satisfy the \emph{trivial assumption} which states that each pattern of $\choreo$ must have $n$ vertices to be performed by a $n$-swarm $\swarm$.
\subsection{The \texttt{Dancing} problems}\label{sec:dancing_problems}
Given a swarm $\swarm$, the general \probfont{Dancing} problem requires $\swarm$ to perform any \emph{feasible} choreography.
A choreography is feasible if it satisfies some necessary conditions so that a swarm of $n$ robots can perform it; indeed, the trivial assumption is one of them.
Moreover, some constraints may be satisfied by the initial configuration of $\swarm$ so that it can perform a feasible choreography.
Formally, we can define a \probfont{Dancing} problem for a swarm $\swarm$ of $n$ robots as the tuple $\mathfrak{D} = \langle\Sigma(n), \mathcal{I}(n), \phi \rangle$, where 
\begin{itemize}
    \item $\Sigma(n)$ defines the \emph{alphabet of patterns}, i.e., the set of patterns with $n$ vertices a feasible choreography can be composed of;
    \item $\mathcal{I}(n)$ defines the set of all configurations from which $\swarm$ can start the choreography;
    \item $\phi$ is a predicate that must be satisfied by all the feasible choreographies (for example, it can limit their length, the presence of repetitions, the periodicity, ...).
\end{itemize}
Thus, $\choreo=(\Pi_0,\dots, \Pi_{q-1})^x$ is a \emph{feasible} choreography if $\Pi_i\in \Sigma(n)$ for any $i\in [0,q-1]$ and $\choreo$ satisfies $\phi$.
Let $Feas(\mathfrak{D})$ be the set of all feasible choreographies for $\mathfrak{D}$.
Note that, for $\mathfrak{D}$ to be well defined, each pattern $\Pi\in \Sigma(n)$ must appear in at least one feasible choreography; in other words, $\phi$ cannot exclude a pattern of the alphabet from all the feasible choreographies.

An algorithm $\algo$ solves $\mathfrak{D}$ if, for any swarm $\swarm$ starting from a configuration in $\mathcal{I}(n)$, it makes $\swarm$ perform any feasible choreography in $Feas(\mathfrak{D})$.

A \probfont{Dancing} problem $\langle\Sigma(n),\mathcal{I}(n), \phi \rangle$ is said to be \probfont{Universal} if $\Sigma(n) =(\Reals^2)^n$ (i.e., any pattern with $n$ vertices) and $\mathcal{I}(n)=(\Reals^2\times \{\off\})^n$ is the set of all possible initial configurations where robots are $\off$-colored.

In this paper, we provide an algorithm under $\LUMI^\seq$ which uses $k$ colors and solves the \probfont{Universal Dancing} problem $\mathfrak{D}_U = ((\Reals^2)^n, (\Reals^2\times \{\off\})^n, \phi)$ where $\phi$ only requires that $q\leq \binom{n+k-7}{k-4}$ (where $q$ is the length/period of a choreography).
\subsection{Colored Counter and Composition Gray Codes}\label{sec:graycodes}
As already mentioned, one of the main benefits of the sequential schedulers is the possibility of implementing a \emph{distributed counter} through the robots' lights.
Here, we describe a method to implement such a counter by exploiting the Gray code.
The size of the counter (i.e., the maximum value that can be counted) increases as the number of available colors and the number of robots in the swarm increase.
Then, in \Cref{sec:algorithm}, we will explain how our algorithm uses this technique to keep track of the ongoing pattern that is being formed, and to reset the counter in case of periodic choreographies.
Thus, the counter size defines a tight bound on the length (if finite) /period (if periodic) $q$ of the choreographies performable by a swarm of $n$ robots with $k$ colors according to our algorithm.

Let $\swarm$ be a swarm of $n$ robots and let $\palette=\{\ell_1,\dots, \ell_k\}$ be the non-empty palette of colors used by $\swarm$.
We assume that $\palette$ is totally ordered: w.l.o.g., let $\ell_1 < \dots < \ell_k$.
We implement a counter whose value iterates over the sequence of integers $0,\dots, z-1$, with $z=\binom{n+k-1}{k-1}$.
Note that $\binom{n+k-1}{k-1}$ corresponds to the number of (weak) compositions of $n$ into $k$ parts\footnote{A weak composition of $n$ into $k$ parts is an ordered representation of $n$ as the sum of $k$ non-negative integers.}. 

Given a configuration $C$ of $\swarm$, we define the \emph{composition vector} (or simply, \emph{counter}) as the $k$-tuple $X = (x_1, x_2, \dots, x_k)$, where $x_i$ denotes the number of robots in $C$ with color $\ell_i$. 
Since each robot has only one color at a time, we have $\sum_{i=1}^{k}x_i = n$, and $0\leq x_i \leq n$.
Let $\compositions{n}{k}$ be the set of all $\binom{n+k-1}{k-1}$ composition vectors made with $n$ robots and $k$ colors. 
If there exists a univocal method to order all the elements in $\compositions{n}{k}$ so that $X_0 < \dots < X_{z-1}$, then the robots can encode the composition vector $X_a$ with the integer $a$, which will be the current value of the swarm counter.
Another desirable property of the order $X_0 < \dots < X_{z-1}$ is that any two consequent vectors $X_{a}=(x_1,\dots, x_k)$ and $X_{a+1}=(x'_1,\dots, x'_k)$ (with $a<z-1$) differ at exactly two indexes: in particular, there exist $i\neq j \in \{1,\dots, k\}$ such that $x_i=x'_i - 1$ while $x_j=x'_j+1$.
This means that if the swarm forms a composition vector $X_a$ and wants to increment the counter by 1, it is sufficient for a $\ell_i$-colored robot to turn its light into $\ell_j$, thus updating the composition vector into $X_{a+1}$.

To implement the order of $\compositions{n}{k}$ that fits our purposes, we use \emph{Gray code} for compositions as described by Klingsberg~\cite{Kling82}.
In this work, the authors present a fast algorithm to generate the Gray codes\footnote{Gray code (patented in 1953 by Frank Gray \cite{Gray53}) is a binary encoding method with the additional property that the representations of two consecutive integers differ by only one bit.} to represent the compositions of $n$ into $k$ parts.
In particular, the algorithm produces the ordered list $\L(n,k)=\langle X_0,\dots,X_{z-1}\rangle$ of all the Gray codes of all the $z=\binom{n+k-1}{k-1}$ compositions.
We illustrate the recursive method to generate the list of gray codes $\L(j,k)$ for any values $j \geq 0$ and $k\geq 2$.
The reverse of the list is denoted by $-\L(j,k)$.
For the base case, $k=2$, we have $\L(j,2) = \langle X_0, \dots , X_j\rangle$ where $X_a = (j-a,a)$.
For an arbitrary $k > 2$, the gray code can be obtained by recursion as follows,
\begin{align*}\label{eq:graycode}
    \L(j, k + 1) = \oplus_{l=0}^j (-1)^l[\L(j-l, k) \otimes \{l\}] 
\end{align*}
where $\oplus$ is the concatenation operator, and $\otimes$ is the Cartesian product of the sets.
From the results of Klingsberg~\cite{Kling82}, we know that this Gray code construction follows the same property: it begins with the composition vector $X_0 = (n, 0,\dots, 0)$ and ends with $X_{z-1} = (0, \dots, 0, n)$.
See \Cref{fig:graycode} in \Cref{appendix:extra} to have some examples.
\subsection{Bounded universal choreographies}\label{sec:bounded_choreographies}
In \Cref{sec:dancing_problems}, we have defined the \probfont{Universal Dancing} problem under study as $\langle (\Reals^2)^n,((\Reals^2) \times \{\off\})^n, \phi \rangle$ where $\phi$ only defines a bound on the length/period $q$ of the feasible choreographies.
We here prove that if we drop such a bound, i.e., if $\phi$ is a tautology $\top$, then the problem is unsolved under our model. Formally
%
\begin{theorem}
    The $\langle (\Reals^2)^n,((\Reals^2) \times \{\off\})^n, \top \rangle$ problem cannot be solved under $\LUMI^\seq$ even if robots have rigid movements and share a global coordinate system. 
\end{theorem}
\begin{proof}
    By contradiction, let $\algo$ be an algorithm for $\langle (\Reals^2)^n,((\Reals^2) \times \{\off\})^n, \top \rangle$ under $\LUMI^\seq$ with $k>1$ colors.
    Suppose $\swarm$ is a swarm of $n$ robots, and let $z=\binom{n+k-1}{k-1}$ be the number of color-compositions of the swarm.
    Let $\choreo = (\mathtt{P}_1, \Pi_1 ,\dots, \mathtt{P}_{z},\Pi_{z}, \mathtt{P}_{z+1}, \Pi_{z+1})$ be a feasible choreography for $\swarm$, such that $\mathtt{P}_i\in \Point$ for any $i\in[1,z+1]$ and $\Pi_i\neq \Pi_j$ for any $i\neq j \in[1,z+1]$.
    In other words, $\swarm$ is asked to perform a choreography that requires the swarm to gather into a point before forming each pattern of the series $\Pi_1, \dots \Pi_{z+1}$.
    Following the definition of choreography, we know that $\Pi_i\notin \Point$.
    Consider an execution of $\algo$ under a \emph{round-robin} activation scheduler $\AS$ (thus, a {\seq} scheduler where each epoch lasts $n$ rounds and robots are always activated in the same order).
    Let $t_b$ be the first time when the swarm forms a point $\mathtt{P}_b$ where robots have the same color-composition as while forming a previous point $\mathtt{P}_a$, with $a<b$.
    Note that $t_b$ exists since the maximum number of $k$-colorings for a swarm of $n$ robots is $z$.
    Let $t_a$ be the time when $\mathtt{P}_a$ was formed under $\AS$.
    W.l.o.g. and for simplicity, let us assume that the same colors in $\mathtt{P}_a$ are assigned to the same robots in $\mathtt{P}_b$.

    Consider the {\seq} scheduler $\AS' = \AS(0)\dots \AS(t_a) \left( \AS(t_{a+1})\dots \AS(t_{b})\right)^\infty$ such that it equals $\AS$ until $t_b$ and then activates the same sequence of robots as starting from $t_a$.
    It is easy to prove that $\AS'$ guarantees the fairness condition too: in fact, all the robots are activated at least once during the time $[t_{a+1}, t_b]$ in $\AS$.
    This is true since, in that time, the swarm passes from a $\Point$ pattern (i.e., $\mathtt{P}_a$), to a non-point pattern (i.e., $\Pi_a$), and lastly it gathers again into a point (i.e., $\mathtt{P}_b$): to perform this sequence of patterns, at least one epoch is needed.
    
    So, if we consider the same problem under $\AS'$, then, after time $t_b$, the robots will perform the same actions as in the interval $[t_{a+1},t_b]$ in a loop.
    This results in the swarm never completing the end of the choreography $\Pi_b,\mathtt{P}_{b+1},\dots,\mathtt{P}_{z+1}, \Pi_{z+1}$, thus never terminating.
    Contradiction achieved.  
\end{proof}
\subsection{Multivertices ranking}\label{sec:multivertices_ranking}
Let $\Pi=\{({v}_1,\card_1), \dots, ({v}_m,\card_m)\}\in\NPoints$ be a pattern with $n$ vertices (and $m$ multivertices), given according to a coordinate system ${\COO}$.
To lighten the notation, we often indicate a multivertex $(v_i,\card_i)$ as only $v_i$.
Let $\Omega$ be the $SEC(\Pi)$ whose center is $O$.
Let us suppose, w.l.o.g., that the positions of the multivertices in $\Pi$ are given considering a coordinate system whose origin and unit distance are $O=(0,0)$ and $radius(\Omega)=1$, respectively.
Let $\Pi_O$ denote $\Pi$ without the multivertex lying in $O$, if it exists, and let $m' = |shape(\Pi_O)|\in\{m,m-1\}$.

We now describe how to obtain an unambiguous rank of the multivertices of $\Pi$ (refer to \Cref{fig:ranking_multivertices} for an example).
For any multivertex $v_i$, we define the following elements:
\begin{itemize}
    \item $\Psi(v_i)$ as the radial projection of $v_i$ on $\Omega$, if $v_i\in\Pi_O$ (i.e., the intersection between the radius where $v_i$ lies and $\Omega$).
    Let $\{\psi_1, \dots, \psi_{m''}\}$ be the set of all the radial projections, where $m''\leq m'$, assuming w.l.o.g. that $[\psi_1, \dots, \psi_{m''}]$ is the cyclic order\footnote{We here consider indexes in the circular range $[1,\cdots, m'']$.} according to the clockwise orientation of $\COO$.
    \item $\theta_i=\angle \psi_j O\psi_{j+1}$ and $\hat{\theta}_i = \angle \psi_jO\psi_{j-1}$ where $\psi_j=\Psi(v_i)$.
        If $v_i$ lies on $O$, we set $\theta_i=\hat{\theta}_i=0$.
    \item $\mu(v_i) = (\rho(v_i),\theta_i,\card_i)$ and $\hat{\mu}(v_i) = (\rho(v_i),\hat{\theta}_i,\card_i)$ where $\rho(v_i) = 1 -\dist{v_i,O}$.
\end{itemize}
The multivertices of $\Pi$ belong to concentric circles of $\Omega$.
Then, it is possible to unambiguously define two cyclic orders on $\Pi_O$: one clockwise according to $\COO$, and the other counterclockwise.
Let $\mathcal{A}=[v_{a_1},\dots, v_{a_{m'}}]$ be the clockwise one, and let $\mathcal{B} = [v_{b_1},\dots, v_{b_{m'}}]$ be the counterclockwise\footnote{$a_1,\dots,a_{m'}$ and $b_1,\dots,b_{m'}$ represent two permutations of the $m'$ indexes of the multivertices of $\Pi_O$.}.
Starting from each $v_{a_i}\in\mathcal{A}$ we define the string $CW(v_{a_i})= \mu(v_{a_i})\mu(v_{a_{i+1}})\dots \mu(v_{a_{i-1}})$ (clockwise);
from each $v_{b_i}\in\mathcal{B}$ we define the string $CCW(v_{b_i})= \hat{\mu}(v_{b_{i}})\hat{\mu}(v_{b_{i+1}})\dots \hat{\mu}(v_{b_{i-1}})$ (counterclockwise).
Let $\gamma(v_i) = \min\{CW(v_i), CCW(v_{i})\}$, where the total order $\leq$ of the strings is given by considering the three numerical orders (distances, angles, and multiplicities) in the triples\footnote{Indeed, any $v_i\in\Pi_O$ corresponds to some $v_{a_j}\in\mathcal{A}$ and some $v_{b_h}\in\mathcal{B}$.}.
If $v_i$ lies on $O$, we set $\gamma(v_i) = (\rho(v_i), \theta_i,\card_i)$ which is $(1,0,\card_i)$ by definition (thus, in this case, string $\gamma(v_i)$ is the concatenation of just one $\mu(\cdot)$).
Formally, given two triples $(\rho,\theta,\card)$ and $(\rho', \theta', \card')$ taken from the domain $[0,1] \times [0,2\pi)\times[1,n]$, we define the following relation:
\begin{equation*}
    (\rho,\theta,\card) \leq (\rho', \theta', \card')  \iff 
    \begin{cases}
        \rho < \rho' &\text{ or}\\
        \rho = \rho' \land \theta < \theta' &\text{ or}\\
        \rho = \rho' \land \theta =  \theta' \land \card \leq \card'\,&
    \end{cases}
\end{equation*}
Thus, the multivertices of $\Pi$ can be partitioned in equivalence classes $\{[v_1]_\equiv, \dots, [v_s]_\equiv\}$ so that $1\leq s\leq m$ and $v_j\in[v_i]_\equiv$ iff $\gamma(v_j) = \gamma(v_i)$.
Indeed, it is possible to define a total strict order (\emph{ranking}) among the classes $[v_{\iota_1}]_\equiv < \dots < [v_{\iota_s}]_\equiv$ which is obtained\footnote{$(\iota_1,\dots,\iota_s)$ represents a permutation of $\{1,\dots, s\}$.} by considering the corresponding strings $\gamma(v_{\iota_j})$ for any $1\leq j \leq s$.   
Note that, according to the definition of $\leq$, the multivertices in $[v_{\iota_1}]_\equiv$ lie on $\Omega$ (in fact, they have the minimal value of $\rho(\cdot)$).
Moreover, if a multivertex lies on $\centersec{\Pi}$, then $[v_{\iota_s}]_\equiv$ is a singleton containing only such a multivertex.

Let $\Pi'\in \Pats{\Pi}$ be a similar pattern to $\Pi$, and let $\tau:\Pi\to \Pi'$ be the related similarity transformation.
It is obvious that, if $\Pi/_\equiv=\{[v_{1}], \dots, [v_{s}]\}$, then $\Pi'/_{\equiv}=\{[\tau(v_{1})], \dots, [\tau(v_{s})]\}$. 
Moreover, the ranking in $\Pi$ is preserved in $\Pi'$, i.e., $[\tau(v_{\iota_1})]_\equiv < \dots < [\tau(v_{\iota_s})]_\equiv$.
\subsection{Chiral angle}\label{sec:chiral_angle}
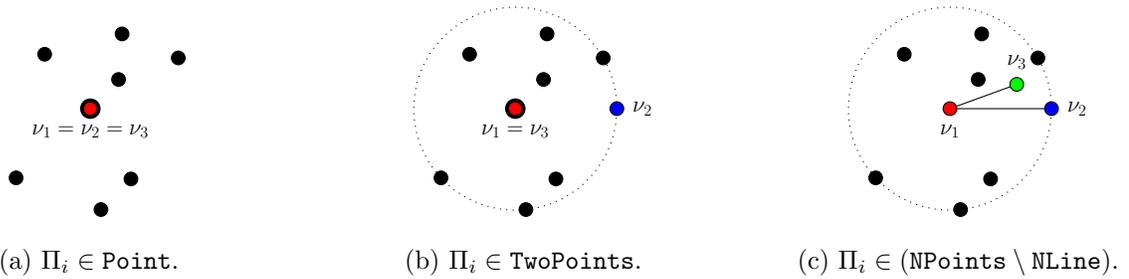
\begin{figure}[h!]
    \centering
    \begin{subfigure}[t]{0.3\textwidth}
    \centering
        \begin{tikzpicture}[scale=0.5, transform shape, font = {\LARGE}]
		\def\r{2.7cm}
        
            \draw [thin, dotted,white] (0,0) circle (\r*1.2);
            
            \mnode{\colorLone}{below}{$\nu_1=\nu_2=\nu_3$}{0,0};
            
            \foreach \a/\b in {30/1,46/0.4,67/0.8,130/0.7,276/1,223/1,300/0.8}
                {\rnode{\colorR}{above}{}{\a:\r*\b};}

	\end{tikzpicture}
        \caption{$\Pi_i\in\Point$.}
        \label{subfig:chiral_angle_Point}
    \end{subfigure}
    \hfill
    \begin{subfigure}[t]{0.3\textwidth}
    \centering
        \begin{tikzpicture}[scale=0.5, transform shape, font = {\LARGE}]
		\def\r{2.7cm}
          
            \draw [thin, dotted] (0,0) circle (\r);
            \draw [thin, dotted,white] (0,0) circle (\r*1.2);

            \mnode{\colorLone}{below}{$\nu_1=\nu_3$}{0,0};
            \rnode{\colorLtwo}{right}{$\nu_2$}{0:\r};
            
            \foreach \a/\b in {30/1,46/0.4,67/0.8,130/0.7,276/1,223/1,300/0.8}
                {\rnode{\colorR}{above}{}{\a:\r*\b};}
            
	\end{tikzpicture}
        \caption{$\Pi_i\in\TwoPoints$. }
        \label{subfig:chiral_angle_TwoPoints_from_Npoints}
    \end{subfigure}
   \hfill
    \begin{subfigure}[t]{0.3\textwidth}
    \centering
        \begin{tikzpicture}[scale=0.5, transform shape, font = {\LARGE}]
		\def\r{2.7cm}

            \draw [thin, dotted,white] (0,0) circle (\r*1.2);
            \draw [thin, dotted] (0,0) circle (\r);
            
            \draw[] (0,0) -- (0:\r);
            \draw[] (0,0) -- (20:\r*0.7);
            \rnode{\colorLone}{below}{$\nu_1$}{0,0};
            \rnode{\colorLtwo}{right}{$\nu_2$}{0:\r};
            \rnode{\colorLthree}{above}{$\nu_3$}{20:\r*0.7};
            \foreach \a/\b in {30/1,46/0.4,67/0.8,130/0.7,276/1,223/1,300/0.8}
                {\rnode{\colorR}{above}{}{\a:\r*\b};}
            
	\end{tikzpicture}
        \caption{$\Pi_i\in(\NPoints\setminus \NLine)$.}
        \label{subfig:chiral_angle_NPoints}
    \end{subfigure}
    \caption{Chiral angles and supporting circles in different scenarios where $C_{|\Reals^2}\in\NPoints$.}
    \label{fig:chiral_angle}
\end{figure}
Let $\choreo=(\Pi_0,\dots,\Pi_{q-1})^x$ be a choreography $\swarm$ must perform starting from an arbitrary initial configuration where all robots are $\off$-colored.
Each robot receives $\choreo$ so that the vertices' coordinates of the patterns are given through a coordinate system, which may be different from robot to robot.
Thus, before making $\swarm$ form a given pattern $\Pi_i$, our algorithm includes a preliminary phase where three robots (called \emph{leaders}) arrange themselves in a triangle $\triangle\nu_1\nu_2\nu_3$ (possibly degenerate), called \emph{chiral angle}.
This triangle provides the swarm a partial (if the triangle is degenerate) or total (if the triangle is not degenerate) global coordinate system of the plane.
Notably, $\nu_1$ represents its origin; if $\nu_1\neq \nu_2$, then $\nu_2$ represents the coordinate $(0,1)$ (thus fixing the unit distance and the $x$-axis position and orientation); if $\nu_3$ is not aligned with $\nu_1$ and $\nu_2$, then $\nu_3$ represents a $y$-positive coordinate (thus fixing the $y$-axis position and orientation, and the clockwise orientation of the plane).

Let $C$ be a configuration of the swarm $\swarm$ where the three leaders are colored as $\Lone$, $\Ltwo$, and $\Lthree$, and let $\Pi_i$ be the pattern of $\choreo$ the swarm must form starting from $C$.
A \emph{chiral angle} for $C$ and $\Pi_i$ is a set of three points $\nu_1,\nu_2,\nu_3$ (possibly coincident) on the plane that unequivocally defines the position, orientation, and scale of the actual pattern ${\Pi}\in\Pats{\Pi_i}$ that $\swarm$ will form.
The three leaders $\Lone$, $\Ltwo$, and $\Lthree$ are intended to arrange on the points $\nu_1,\nu_2,\nu_3$, respectively.

Once $\Lone$ and $\Ltwo$ have fixed $\nu_1$ and $\nu_2$ for $\Pi_i$, all the robots can construct the circle $\Omega_i$ whose center is $\nu_1$ and whose radius is $\dist{\nu_1,\nu_2}$.
We call $\Omega_i$ as the \emph{supporting circle} of $\Pi_i$.
If $\Pi_i\in(\Point\cup\NPoints)$, then the swarm must form $\Pi$ so that $\Omega_i=SEC(\Pi)$ (trivially, it means that if $\Pi_i\in\Point$, then $\Omega_i$ will be a degenerate circle).
Otherwise, if $\Pi_i\in(\TwoPoints\cup \ThreePoints)$, then the swarm must form $\Pi$ so that $\overline{\nu_1\nu_2}$ is the (longest) edge of $\Pi$.
Note that, in all the cases, all the vertices of $\Pi$ are contained within $\Omega_i$.

Note that our strategy for constructing the supporting circles is to guarantee a spatial homogeneity in the $\choreo$ performance: notably, all the supporting circles are concentric.
In fact, regardless of $\Pi_i$, $\nu_1$ is static at the same point of the plane, which corresponds to $\centersec{C}$ if $C_{|\Reals^2}\in\NPoints$, otherwise $\nu_1$ corresponds to the position of $\Lone$, denoted by $pos(\Lone)$ (thus, $\Lone$ does not move since it already lies on $\nu_1$).
Moreover, given any subsequence $\Pi_a,\dots, \Pi_b$ of $\choreo$, if $\Pi_i\notin\Point$ for any $i\in [a,b]$, then $\Omega_{a},\dots,\Omega_{b}$ are equal.

We now explain how to compute $\nu_2$ and $\nu_3$ for the different cases, and the actual pattern $\Pi\in\Pats{\Pi_i}$ to be formed:

If $\Pi_i\in\Point$, the chiral angle (and thus $\Omega_i$) degenerates on the same point $\nu_1=\nu_2=\nu_3$, which is the vertex at which the swarm will gather (see \Cref{subfig:chiral_angle_Point}).
Formally, $\Pi=\{(\nu_1,n)\}$.

If $\Pi_i\in\TwoPoints$, then $\nu_3=\nu_1$, while $\nu_2$ is a different point: the swarm will arrange itself on these two vertices with the respective multiplicities.
The point $\nu_2$ is computed in this way: if $\Ltwo$ does not lie on $\nu_1$, then its position becomes the new $\nu_2$ for $\Pi_i$ (see \Cref{subfig:chiral_angle_TwoPoints_from_Npoints}); otherwise, the position $(0,1)$ according to the local coordinate system of $\Ltwo$ becomes the new $\nu_2$.
Formally, if $\Pi_i=\{(v_1,\card_1), (v_2, \card_2)\}$ with $\card_1\geq \card_2$, then $\Pi=\{(\nu_1,\card_1), (\nu_2,\card_2)\}$.

If $\Pi_i\in\ThreePoints$, then $\nu_1,\nu_2,\nu_3$ are three distinct points forming the target triangle on which the swarm must arrange.
Point $\nu_2$ is computed as before, and it fixes the longest edge $\overline{\nu_1\nu_2}$ of the target pattern $\Pi$.
Once $\Lone$ and $\Ltwo$ have fixed $\nu_1$ and $\nu_2$, then $\Lthree$ chooses the position\footnote{Note in fact that different positions of $\nu_3$ can be found in order to form the target triangle.} of $\nu_3$, so that $\triangle\nu_1\nu_2\nu_3\in \Pats{\Pi_i}$ and $\dist{\nu_1,\nu_2}\geq \dist{\nu_1,\nu_3}\geq \dist{\nu_2,\nu_3}$.
Formally, if $\Pi_i=\{(v_1,\card_1), (v_2, \card_2), (v_3,\card_3)\}$ with $\dist{v_1,v_2}\geq \dist{v_1,v_3}\geq \dist{v_2,v_3}$, with $\card_1\geq \card_2$, and with $\card_1\geq \card_3$ if $\dist{v_1,v_3}=\dist{v_1,v_2}$, then $\Pi=\{(\nu_1,\card_1), (\nu_2,\card_2), (\nu_3, \card_3)\}$.

If $\Pi_i\in\NPoints$, then $\nu_1,\nu_2,\nu_3$ are three distinct points forming a triangle with $\dist{\nu_1,\nu_2}\geq \dist{\nu_1,\nu_3}$.
Even in this case, $\nu_2$ is computed as for the $\TwoPoints$ case.
The swarm must form $\Pi$ so that $SEC(\Pi)=\Omega_i$ (which is defined by $\nu_1$ and $\nu_2$), so that $\nu_2$ corresponds to one vertex of $\Pi$.
The precise position of $\Pi$ within $\Omega_i$ will be given by $\Lthree$ once set on $\nu_3$.
Let us describe how $\Lthree$ computes $\nu_3$.
Assume that the ranking of the multivertices of $\Pi_i$ is $[v_{\iota_1}]_\equiv <\dots <[v_{\iota_s}]_\equiv$.
Then $\nu_2$ will be one multivertex of $\Pi$ corresponding to one multivertex of $\Pi_i$ in $[v_{\iota_1}]$.
As seen in \Cref{sec:multivertices_ranking}, we know that the vertices in the minimal class $[v_{\iota_1}]$ lie on $SEC(\Pi_i)$.
So, $\Lthree$ chooses one vertex in $[v_{\iota_1}]$, say w.l.o.g. $v_{\iota_1}$, and then it computes a similarity transformation $\tau$ such that $\tau(v_{\iota_1})=\nu_2$.
If $\Pi_i\in\NLine$, then $\Lthree$ chooses the multivertex closest to $\nu_2$ but not lying on $\nu_1$, and moves there.
Otherwise, $\Lthree$ chooses the multivertex closest to $\nu_2$ but not collinear with $\nu_1$ and $\nu_2$: if multiple multivertices are eligible, then $\Lthree$ chooses one belonging to the lowest class in the ranking (see \Cref{subfig:chiral_angle_NPoints}).
This strategy allows all robots to unambiguously reconstruct $\tau(\Pi_i) = \Pi$.
\Cref{tab:chiral_angle} in \Cref{appendix:extra} summarizes how $\triangle \nu_1\nu_2\nu_3$ is computed.
\subsection{Robot-vertex matching with Dyck words}
Our algorithm adopts a technique to match each robot with its target vertex by exploiting the Dyck words on the Boolean alphabet.
Dyck words are well-parenthesized strings over an alphabet of brackets \cite{Chomsky63,Ginsburg66}; in our case, 0 is the opening bracket, while 1 is the closing one.

Given a boolean string $x$, $|x|$ denotes its length, and $|x|_0$ ($|x|_1$, resp.) denotes the number of 0s (1s, resp.) contained in $x$.
A boolean string $x$ is \emph{balanced} if $|x|_0=|x|_1$.
The empty string (i.e., without symbols) is denoted by $\epsilon$.
For any $i\in [0,|x|]$, we indicate with $\delta_{x}(i) = |y|_0 - |y|_1$ where $y$ is a prefix of $x$ of length $i$.

We consider the Dyck language $\Dyck = \{x\in \{0,1\}^* \sucht |x|_0=|x|_1 \land (\delta_{x}(i)\geq 0 \; \forall i \in [0,|x|]) \}$.
In other terms, $x$ is a Dyck word if it is balanced and any prefix of $x$ must not have more 1s than 0s.
Equivalently, $x\in\Dyck$ if \emph{(i)} $x=\epsilon$, or \emph{(ii)} $x=0y1z$ for some $y,z\in\Dyck$ where the 0 and 1 that envelops $y$ are a pair of \emph{matched brackets}.
If $x=x_1\cdots x_m$ is a Dyck word, then $x_b=1$ is matched with $x_a=0$ where $a=\max\{1\leq i<b \sucht x_i=0 \land \delta_x(i)-1 = \delta_x(b)\}$.
Condition \emph{(ii)} implies that $\Dyck$ is closed under concatenation.

Given a balanced string $x$, we say that $x$ is \emph{minimal} if it cannot be non-trivially factorized into multiple balanced strings. 
E.g., $110100$ is minimal, while $010011$ is not minimal since it can be factorized into $01$ and $0011$.
    \section{Algorithm}\label{sec:algorithm}
\subsection{Algorithm Outline}
In this section, we present our algorithm solving the \probfont{Universal Dancing} problem $\mathfrak{D}_U$ presented in \Cref{sec:dancing_problems} for a swarm $\swarm=\{r_1,\dots,r_n\}$ of $n\geq 3$ robots that satisfies all the hypotheses presented in \Cref{sec:preliminaries}.
In particular, we exhibit an algorithm with $k\geq 4$ colors which allows $\swarm$ to perform any feasible choreography $\choreo=(\Pi_0,\dots,\Pi_{q-1})^x$ with $x\in\{1,\infty\}$ that satisfies $q\leq \binom{n+k-7}{k-4}$.

Our algorithm uses a palette $\palette=\{\ell_1,\dots, \ell_{k-3},\Lone,\Ltwo,\Lthree\}$.
By convention, we assume that $\ell_1 = \colr{off}$, which is the default color to which all robots are initialized.
Notably, $\Lone$, $\Ltwo$, and $\Lthree$ are dedicated to the three leader robots; instead, the other $k-3$ colors are used by the $n-3$ non-leader robots for implementing the counter and thus to keep track of the pattern to be formed in $\choreo$.
We assume that $\palette$ is totally ordered, e.g., $\ell_1< \dots < \ell_{k-3} < \Lone<\Ltwo<\Lthree$.
Let the current counter vector be $X_i = (x_1, \dots, x_{k-3})$, where $x_j$ represents the number of robots of color $\ell_j\in \palette\setminus\{\Lone, \Ltwo, \Lthree\}$, for $j\in[1, k-3]$. 
This means that $\Pi_i$ is the ongoing pattern to be formed in the choreography.

The core of our algorithm is composed of three phases, i.e., \Phase{1}, \Phase{2}, \Phase{3}, which are repeated cyclically until the choreography ends.
For each $\Pi_i\in\choreo$, \Phase{1} aims at setting the chiral angle; \Phase{2} aims at making robots form $\Pi_i$; \Phase{3} aims at updating the counter value.
Indeed, if the choreography is periodic, then the algorithm execution never ends.
Before entering into the loop of the three phases, we need an initial phase \Phase{0} to set up the three leader robots.
See \Cref{algo:pseudocode} in \Cref{appendix:extra} for the pseudo-code of the whole algorithm.
\subsection{\Phase{0} - Election of leaders}
This phase starts from the initial configuration $C$ (where all the robots are $\off$-colored) and terminates when the three leaders have elected themselves by setting the dedicated colors.
No robot moves during \Phase{0}.
If $C_{|\Reals^2}\notin\ThreePoints$, the first activated $\off$ robot lying on $SEC(C)$ sets its color to $\Ltwo$.
Otherwise, a $\off$ robot lying on one of the longest edges of the triangle sets to $\Ltwo$.
Let us see which $\off$ robot becomes $\Lone$: if $C_{|\Reals^2}\in \Point$, the next activated $\off$ robot sets its color to $\Lone$.
If $C_{|\Reals^2}\in (\TwoPoints\cup\ThreePoints)$, the $\off$ robot lying on the other endpoint of the (longest) edge of $C_{|\Reals^2}$ sets to $\Lone$, so that $pos(\Lone) \neq pos(\Ltwo)$.
Otherwise, the $\off$ robot closest to $\centersec{C}$ on a distinct position from $pos(\Ltwo)$ becomes $\Lone$.
Note that, in this phase, $\Ltwo$ always lies on $SEC(C)$, while $pos(\Lone)= pos(\Ltwo)$ iff $C_{|\Reals^2}\in \Point$.

Lastly, the next activated $\off$ robot turns its color to $\Lthree$, thus completing the three leaders' setting.
When no ambiguity arises, we denote with $\Lone$, $\Ltwo,$ and $\Lthree$ both the colors and the related leader robots.
All the other $n-3$ non-leaders are $\off$, thus the counter value is 0: now the loop can start with the formation of $\Pi_0$ (i.e., the first pattern of the choreography).
\subsection{\Phase{1} - Chiral angle setup}
Let $C$ be the current configuration.
Let $i$ be the counter's value, and let $\Pi_i$ be the corresponding pattern in the choreography.
\Phase{1} starts when $\Pi_i$ is not formed by the swarm and the three leaders are not located on the chiral angle related to $C$ and $\Pi_i$.
In this phase, the three leaders $\Lone$, $\Ltwo$, $\Lthree$ compute the chiral angle $\triangle\nu_1\nu_2\nu_3$ as explained in \Cref{sec:chiral_angle}, and sequentially reach the relative vertices.
Specifically, $\Lone$ moves to $\centersec{C}$ if $C_{|\Reals^2}\in \NPoints$, otherwise it stays still since it is already on $\nu_1$.
Then, $\Ltwo$ calculates the position of $\nu_2$ for $\Pi_i$, and moves to $\nu_2$.
Note that, if $C_{|\Reals^2}\notin\Point$, $\nu_2=pos(\Ltwo)$; otherwise, all the robots lies in $(0,0)$ (since all have an egocentric coordinate system) and $\Ltwo$ is required to move in a distinct position, conventionally to $(0,1)$ (thus, the request is satisfied even if $\Ltwo$ is stopped before reaching its point $(0,1)$). 
Lastly, $\Lthree$ calculates and moves to $\nu_3$, completing the chiral angle $\triangle \nu_1\nu_2\nu_3$ for $\Pi_i$.
No other robot moves in this phase.
\subsection{\Phase{2} - Pattern Formation}
This phase starts when the leaders have formed the chiral angle for $\Pi_i$ and ends as soon as the swarm forms the pattern $\Pi\in\Pats{\Pi_i}$ which is unequivocally defined by the chiral angle.
Let $r$ be an activated non-leader robot.
Thanks to the counter's value of the swarm, $r$ detects the pattern $\Pi_i$ to be formed.
If $\Pi_i$ is not formed and the triangle $\triangle \Lone\Ltwo\Lthree$ constitutes a valid chiral angle for $\Pi_i$, then $r$ knows it is in \Phase{2}.
Thus, $r$ detects the actual pattern $\Pi$ on whose vertices robots must arrange themselves. 
Let us distinguish the two cases:
\subsubsection{Case $\Pi\notin \NPoints$}
If $\Pi\in\Point$ (thus, $\nu_1=\nu_2=\nu_3$), then $r$ reaches the gathering point where the leaders lie.
If $\Pi\in(\TwoPoints\cup \ThreePoints)$, then if $r$ lies on a (un)saturated multivertex of $\Pi$, it stays still.
Otherwise, it reaches the closest unsaturated multivertex of $\Pi$.
The leaders do not move, since they already lie on their target vertices.
\subsubsection{Case $\Pi\in\NPoints$}
Let $C$ be the current configuration, where $\Lone$, $\Ltwo$, and $\Lthree$ form a non-degenerate chiral angle.
Let $\Omega$ be the supporting circle for $\Pi$, centered in $\Lone$ and such that $\Ltwo$ belongs to its boundary.
We remind that, in this case, $\Ltwo$ and $\Lthree$ already lie on their target vertices of $\Pi$ by construction, thus they will not move during $\Phase{2}$.
In this phase, all non-leaders reach their target vertices of $\Pi$, and subsequently, $\Lone$ possibly moves to the last missing vertex left free.
Note that $\Lone$ already lies on its target vertex (and thus it will not move in this phase) only if $\nu_1$ is a vertex of $\Pi$.
If $\nu_1\notin\Pi$, then, robots must properly agree on the target (multi)vertex to be left free for $\Lone$: notably, they select the closest multivertex to $\Lone$ (in case of equidistance, the orientation given by the chiral angle unambiguously selects a multivertex).

Let $D$ be the \emph{main diameter}, i.e., the diameter of $\Omega$ starting from $\Ltwo$.
Given a vertex $v$ of $\Pi$, we indicate with $v_{\perp}$ its projection on $D$.
We call the segment $\overline{vv_\perp}$ the \emph{line projection} of $v$.
Let $\Pi_\perp = \multiset{v_\perp \sucht v\in \Pi}$.

This case is handled in four sub-phases: \emph{(i)} all non-leaders move to $D$ perpendicularly, \emph{(ii)} they shift along $D$ to arrange on the vertices of $\Pi_\perp$ (except for the three vertices which are intended for the leaders), \emph{(iii)} they move perpendicularly w.r.t. $D$ to reach the corresponding target vertices of $\Pi$, and \emph{(iv)} $\Lone$ reaches the last missing vertex of $\Pi$.

Note that this strategy (i.e., projecting $\Pi$ on $D$ and making robots arrange on $\Pi_\perp$ before reaching $\Pi$) guarantees a constant bound on the maximum distance traveled by each robot so that the number of epochs remains independent of the number of robots, even under non-rigid movements.
Referring to \Cref{fig:phase2}, we explain the four sub-phases in detail:
\begin{figure}[t]
    \centering
    \def\crosssize{3pt}
    \tikzset{
        font={\LARGE},
        every picture/.style={scale=0.65,transform shape}
    }

    \def\robotcoordsOriginal{30/1,46/0.4,67/0.8,130/0.7,276/1,223/1,300/0.8}
    \def\robotcoordsExtra{20/0.7}
    \def\robotcoords{\robotcoordsOriginal}

    \def\crosscoords{15/1,60/0.6,95/0.9,145/1,255/0.9,235/1,315/0.9, 0/1, 20/0.7}

    \begin{subfigure}[t]{0.48\textwidth}
        \centering
        \begin{tikzpicture}[scale=0.6, transform shape, font = {\LARGE}]
            \def\r{4cm}
            \draw[thin] (-\r,0)--(\r,0);
            \draw[thin,dotted] (0,0) circle(\r);

            \mnode{\colorLone}{below}{$\Lone$}{0,0};
            \mnode{\colorLtwo}{right}{$\Ltwo$}{\r,0};
            \mnode{\colorLthree}{above}{$\Lthree$}{20:0.7*\r};

            \foreach[count=\i] \a/\b in \robotcoordsOriginal {%
                \draw [dashed,gray!60] (\a:\r*\b) -- (0:{\r*\b *cos(\a)});
            }
            \foreach[count=\i] \a/\b in \robotcoordsOriginal {%
                \rnode{\colorR}{above}{}{\a:\r*\b};
            }
        \end{tikzpicture}
        \caption{Non-leader robots and their projections on $D$.}
        \label{subfig:phase2_migration_D}
    \end{subfigure}
    \hfill
    \begin{subfigure}[t]{0.48\textwidth}
        \centering
        \begin{tikzpicture}[scale=0.6, transform shape, font = {\LARGE}]
            \def\r{4cm}
            \draw[thin] (-\r,0)--(\r,0);
            \draw[thin,dotted] (0,0) circle (\r);

            \foreach[count=\j] \a/\b in \crosscoords {%
                \draw [dashed,gray!60] (\a:\r*\b) -- (0:{\r*\b *cos(\a)});
            }
            \foreach[count=\j] \a/\b in \crosscoords {%
                \draw[thick,gray!80] 
                ([shift={(0:\r*\b *cos(\a))}] -\crosssize, -\crosssize) -- ([shift={(0:\r*\b *cos(\a))}] \crosssize, \crosssize)
                  ([shift={(0:\r*\b*cos(\a))}] \crosssize, -\crosssize) -- ([shift={(0:\r*\b*cos(\a))}] -\crosssize, \crosssize);
                \draw[\colorB,thick]
                  ([shift={(\a:\r*\b)}] -\crosssize, -\crosssize) -- ([shift={(\a:\r*\b)}] \crosssize, \crosssize)
                  ([shift={(\a:\r*\b)}] \crosssize, -\crosssize) -- ([shift={(\a:\r*\b)}] -\crosssize, \crosssize);
                
            }
        \end{tikzpicture}
        \caption{Projections of the target vertices for the non-leaders.}
         \label{subfig:phase2_target_points}
    \end{subfigure}

    \caption{\Phase{2}, case $\Pi\in(\NPoints\setminus \NLine)$.}
    \label{fig:phase2}
\end{figure}
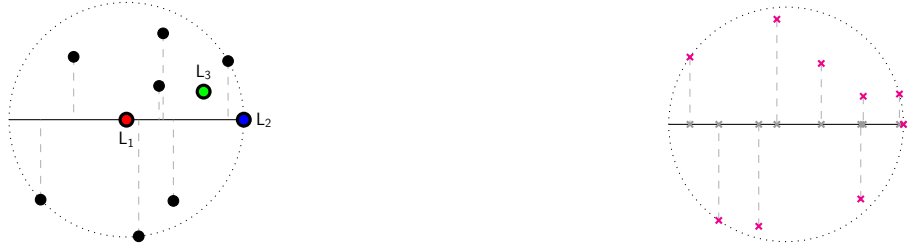
\subparagraph{Migration towards $D$.}
Each non-leader robot reaches $D$ moving along a perpendicular trajectory (see \Cref{subfig:phase2_migration_D})
\subparagraph{Arrangement on $D$.}
Once all $m=n-3$ non-leader robots have reached $D$, they compute $\Pi_\perp$ and remove from it the projections of the three vertices intended for the leaders.
From here on, in this paragraph, we will use robot to refer to a non-leader robot, and projection to refer to a vertex projection intended for a non-leader on $D$.
So, the robots use the following robot-projection matching algorithm to detect the projection they have to reach along $D$.

\begin{figure}[t]
   \centering
        \begin{tikzpicture}[scale=0.7, transform shape, font = {\Large}]
        \def\l{14cm}
        \draw[dotted] (0,0) -- (\l+1,0);
        \coordinate (nu2)   at (0,0);
        \node at (\l*1.05,0)    {$D$};

        \rnode{\colorLtwo}{left}{$\Ltwo$}{nu2};

        \draw[shorten >=0.5,->]   
                    (1,0) edge      [bend right=45]        node {} (2,0)
                    (11,0) edge      [bend right=50]        node {} (12,0)
                    (6,0) edge      [bend left=50]        node {} (5,0)
                    (8,0) edge      [bend left=50]        node {} (7,0)
                    (9,0) edge      [bend left=50]        node {} (4,0)
                    (10,0) edge      [bend left=50]        node {} (3,0)
                    ;
        
        \foreach \x in {2,3,4,5,7,12} {
            \tnode{above}{$0$}{\x,0};
        }
    
        \foreach \x in {1,6,8,9,10,11} {
            \rnode{\colorR}{above}{$1$}{\x,0};
        }

        \rnode{\colorR}{above}{$0/1$}{13,0};
        \tnode{above}{}{13,0};
        
    \end{tikzpicture}
    \caption{Robot-projection matching. In this case, $w=100001011110$ is factorized in $x=10$, $y=00010111$, and $z=10$. $x^R$, $y$, and $z^R$ are Dyck words. The rightmost robot and the vertex where it lies are not included in $w$.}
    \label{fig:dyck_word}
\end{figure}
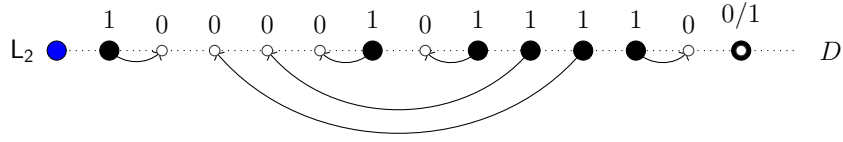

Let $r$ be an activated robot on $D$.
It computes the boolean string $w\in \{0,1\}^{\leq 2m}$ that represents the ordered arrangement of robots and projections along $D$, starting from the closest to $\Ltwo$ (see \Cref{fig:dyck_word}).
In particular, each 1 represents a robot that does not lie on a vertex projection, while each 0 represents a projection not covered by a robot.
Note that multiplicities (multivertices, resp.) on $D$ are treated by unrolling and representing them through factors of adjacent 1s (0s, resp.).
Then, $w$ can be factorized uniquely into minimal-length factors, such that they have the same number of 0s and 1s (see \Cref{lemma:w_factors_dyck_words}).
Let $x=x_1,\dots, x_h$ be such a factor of $w$.
Then, $x$ has this property: $x_1 = 1-x_{h}$ (see \Cref{lemma:strings_first_last_different}).
So, consider $x$ if $x_1=0$, otherwise consider $x^R$ (i.e., its reverse); for simplicity, we will refer to $x$ in both cases.
Thus, $x$ is a Dyck word.
So, if $x_i=1$ corresponds to $r$, then $r$ must reach the projection represented by the corresponding opening 0, say $v_\perp$.

Note that during the movement towards $v_\perp$, $r$ may be stopped while it is crossing a (previously) uncovered projection, say $v'_\perp$ (see \Cref{subfig:dyck_word_r_stops_another_vertex}).
By construction, at the next round, the corresponding 1 and 0 related to $r$ and $v'_\perp$ are removed from $w$, and the matching algorithm can restart from the new arrangement string (see \Cref{subfig:dyck_word_new_matching_another_vertex}).

Otherwise, it may happen that, during its trajectory towards $v_\perp$, $r$ is stopped on a non-vertex point\footnote{Note that this case also comprehends the case when $r$ stops on a multivertex (over)saturated.} (see \Cref{subfig:dyck_word_r_stops_no_vertex}).
In this case, the new arrangement string and thus the new matching can change (see \Cref{subfig:dyck_word_new_matching_no_vertex}).
However, the new matching always ensures that the new intended vertex for $r$ is in the same direction as $v_\perp$ (thus, $r$ does not have to retrace its trajectory backwards).

This sub-phase continues until each of the $m$ vertex projections on $D$ has been covered by a non-leader robot.
\subparagraph{Back to the vertices.}
This sub-phase starts as soon as all the non-leader robots have arranged themselves on the vertex projections on $D$, and ends when all non-leader robots have reached the corresponding vertices traveling perpendicularly w.r.t. $D$.

Let us explain the invariant that each robot $r$ can evaluate to understand that this sub-phase is in progress.
Thanks to the counter (kept by the non-leaders) and the chiral-angle (kept by the leaders), $r$ detects $\Pi_i$ (the pattern of the choreography), the supporting circle $\Omega_i$, and the actual pattern $\Pi$ to be formed (so that $\Omega_i=SEC(\Pi)$). 
If $\Pi$ is formed, then \Phase{2} is terminated.
Otherwise, for each multivertex $(v,\card)$ of $\Pi$, $r$ checks if these two properties hold:
\begin{itemize}
    \item if $(v,\card)$ is not the multivertex left for $\Lone$, $\card$ non-leaders lie on the line projections of $v$;
    \item if $(v,\card)$ is the multivertex left for $\Lone$, $\card-1$ non-leaders lie on the line projections of $v$.
\end{itemize}
Assume $r$ is a non-leader robot lying on $D$, which has positively checked the invariant for this sub-phase.
Then it unambiguously chooses the closest uncovered vertex whose projection corresponds to its current position on $D$, and for which there does not yet exist a robot moving towards it.
Note that multiple vertices (belonging to both the semi-circles cut by $D$) can have the same projections; in this case, $r$ unambiguously elects the target vertex by considering the clockwise orientation of the plane.
After the vertex choice, $r$ starts traveling perpendicularly to $D$ towards the computed vertex.

Robot $r$ may be stopped before reaching its vertex.
If $r$ is a non-leader robot not lying on $D$, then it evaluates the sub-phase invariant.
Let $h$ be the segment perpendicular to $D$ which starts from $D$, contains $r$, and ends at $\Omega_i$.
Let $v_1,\dots,v_s$ be the list of vertices along $h$ so that $\dist{v_i,D} \geq \dist{v_{i+1},D}$.
Then $r$ moves to $v_j$ where $j\in[1,s]$ is the greatest index such that $v_j$ is not covered and $r$ lies on the segment $[v_j,v_{j_\perp})$.
\subparagraph{$\Lone$ towards its vertex.}
Once all the $m-3$ robots have reached their target vertices, the last missing vertex, say $w$, must be covered by $\Lone$.
Let $C$ be the current configuration: remind that $\Lone$ lies on $\centersec{C}$.
By construction, $w$ has been selected for $\Lone$ among the closest to $\centersec{C}$.
If $\centersec{C}=w$, then $\Lone$ does not move since it already covers the last vertex, and the pattern is formed.
Otherwise, $\Lone$ moves to $w$.
Assume $\Lone$ is stopped before reaching $w$, and let $C'$ be the resulting configuration.
Thus, any activated non-leader robot observes that $\Lone$ does not lie on $\centersec{C'}$ and that the swarm does not form $\Pi$: thus, it does nothing.
As soon as $\Lone$ is reactivated, it recomputes $w$ and moves there.
\subsection{\Phase{3} - Counter Update}
Let $C$ be the current configuration and $i$ be the counter value, which points to the pattern $\Pi_i$ of the choreography $\choreo=(\Pi_0,\dots, \Pi_{q-1})^x$.
This phase starts when $C_{|\Reals^2}=\Pi$, where $\Pi\in\Pats{\Pi_i}$ is the pattern defined by the chiral angle.
In $C$, the $n-3$ non-leader robots are colored with a color from $\palette\setminus\{\Lone, \Ltwo, \Lthree\}$.
Let the current counter vector be $X_i = (x_1, \dots, x_{k-3})$, where $x_j$ represents the number of robots of color $\ell_j\in \palette\setminus\{\Lone, \Ltwo, \Lthree\}$, for $j\in[1, k-3]$. 
We distinguish three actions:
\begin{itemize}
    \item\textbf{Termination.} If $i=q-1$ and $x=1$, then the swarm has terminated the choreography, and then it does not update the counter: this causes the swarm to freeze in $\Pi_{q-1}$.
    \item\textbf{Increment.} If $i<q-1$, then the swarm must increment the counter from $i$ to $i+1$.
        Let $X_{i+1}$ be the next vector for counter $i+1$. 
        From the Gray code construction, there are exactly two differences between $X_i$ and $X_{i+1}$ corresponding to one decrement value and one increment value. 
        Let $a$ and $b$ be the two indices that change  between $X_i$ and $X_{i+1}$.
        There must be $x_a$ robots with color $\ell_a$. 
        When a robot with color $\ell_a$ gets activated and realizes that pattern $\Pi_i$ is formed, it changes its color to $\ell_b$. 
        Now the counter vector corresponds to $i+1$, and all robots realize that the pattern $\Pi_{i+1}$ is not formed. 
        \Phase{1} starts again with the movement of $\Lone$, $\Ltwo$, and $\Lthree$ to form the chiral angle. 
    \item\textbf{Resetting.} If $i=q-1$ and $x=\infty$, then the swarm must reset the counter to the initial value $X_0$, i.e., where all the non-leader robots are                 $\colr{off}$-colored.
        However, since a swarm passing from $X_{q-1}$ to $X_0$ under $\seq$ creates other counter vectors and thus may create unambiguous configurations, the swarm must mark this resetting phase through an unequivocal configuration.
        Specifically, we make the leader $\Lthree$ set its color to $\Ltwo$; in this way, the presence of two $\Ltwo$ robots signals non-leader robots that they have to reset the counter.
        Once all non-leader robots have turned into $\colr{off}$, then one of the two $\Ltwo$ robots must turn into $\Lthree$.
        If it is possible to establish which of the two $\Ltwo$ robots was the former $\Lthree$ (e.g., if one $\Ltwo$ does not lie on $SEC(C)$), then leaders maintain their former identities.
\end{itemize}

\subsection{Putting It All Together}
The following theorem summarizes our contribution.
Due to a lack of space, we have moved the correctness and complexity proofs in \Cref{appendix:proofs}.
\begin{theorem}
    The \probfont{Universal Dancing} problem $\mathfrak{D}_U=((\Reals^2)^n,(\Reals^2 \times \{\off\})^n, \phi)$ where $\phi$ only requires that any feasible choreography has length/period $q\leq \binom{n+k-7}{k-4}$ for $k\geq 4$, can be solved under $\LUMI^\seq$ by a swarm of $n$ robots with $k$ colors, even assuming non-rigid movements.   
\end{theorem}

    \section{Conclusions}
In this paper, we have studied the computational power of $\LUMI$ robots under sequential schedulers, concerning their ability to perform sequences of patterns (aka choreographies).
We have proved that their peculiar capability of implementing a distributed counter mechanism allows them to solve the \probfont{Universal Dancing} problem, requiring a swarm to perform (periodic or finite) sequences of any type of pattern, starting from any initial configuration.
We have proved that the number of robots and available colors define a bound on the length/period of the feasible choreographies.
We have presented an algorithm solving \probfont{Universal Dancing} for completely disoriented robots, and which ensures a sort of spatial homogeneity to the performed choreography, even assuming non-rigid movements.

Two possible future research lines can follow this paper: the first can define other classes of \probfont{Dancing} problems, and investigate under which assumptions they are solved. 
Secondly, further work may deeply investigate the distinctive power of robots under sequential schedulers.

    \bibliography{refs}

    \appendix\label{appendix}
    \section{Proofs}\label{appendix:proofs}
We here prove the correctness of our algorithm and we analyze its complexity in terms of runtime and colors.

\subsection{Correctness of \Phase{1}}
\begin{lemma}\label{lemma:pi_unambiguous}
    Given a pattern $\Pi_i$ to be formed, the construction of the chiral angle $\triangle\nu_1\nu_2\nu_3$ as explained in \Cref{sec:chiral_angle} unequivocally defines one pattern in $\Pats{\Pi_i}$.
\end{lemma}
\begin{proof}
    Let us prove the statement for any case:
    \begin{itemize}
        \item If $\Pi_i \in (\Point\cup \TwoPoints)$, the proof follows straightforwardly from the definition of chiral angle. 
        In fact, if $\Pi_i \in \Point$, then $\nu_1=\nu_2=\nu_3$ unequivocally defines the gathering point for the robots.
        If $\Pi_i \in \TwoPoints$, then the chiral angle corresponds to a degenerate triangle with $\nu_1=\nu_3\neq\nu_2$, thus these two points correspond to the gathering points. If the two multivertices of $\Pi_i$ have different cardinalities, then $\nu_1$ corresponds to the multivertex with the greatest cardinality.

        \item If $\Pi_i= \{(v_1,\card_1),(v_2,\card_2),(v_3,\card_3)\}\in \ThreePoints$, then $\triangle\nu_1\nu_2\nu_3\in\Pats{\Pi_i}$ by construction. 
        Let us now prove that the cardinalities of the multivertices of $\Pi_i$ are unequivocally assigned to the points $\nu_1,\nu_2,\nu_3$.
        Assume $\dist{v_1,v_2}\geq \dist{v_1,v_3} \geq \dist{v_2,v_3}$, with $\card_1\geq \card_2$, and with $\card_1\geq \card_3$ if $\dist{v_1,v_3}=\dist{v_1,v_2}$.
        Note that such a ranking of the multivertices of $\Pi_i$ is unambiguous.
        Then $\overline{\nu_1\nu_2}$ corresponds to the longest edge $\overline{v_1v_2}$ where the cardinality of $\nu_1$ will not be less than the cardinality of $\nu_2$.
        If $\Pi_i$ is equilateral (isosceles, resp.), then the cardinality of $\nu_1$ will not be less than the cardinality of $\nu_3$.
        These constraints unequivocally define the pattern $\Pi=\{(\nu_1,\card_1),(\nu_2,\card_2),(\nu_3,\card_3)\}$ with $\dist{\nu_1,\nu_2}\geq \dist{\nu_1,\nu_3}\geq \dist{\nu_2,\nu_3}$.
        
        \item If $\Pi_i=\{(v_1,\card_1),\dots,(v_m,\card_m)\}\in\NPoints$, then $\triangle\nu_1\nu_2\nu_3$ is a triangle with three distinct vertices by construction, with $\dist{\nu_1,\nu_2}\geq \dist{\nu_1,\nu_3}$, and such that $\{\nu_1,\nu_2,\nu_3\}$ are aligned if and only if $\Pi_i\in \NLine$.
        We show that it defines a unique similarity transformation $\tau:\Pi_i\to\Pi$, where thus $\Pi\in\Pats{\Pi_i}$.
        Remember that $\nu_1$ and $\nu_2$ define the supporting circle for $\Pi$, i.e., its smallest enclosing circle, where $\nu_1$ is the center and $\overline{\nu_1\nu_2}$ corresponds to its radius.
        Thus:
        \begin{itemize}
          \item $\nu_1$ fixes the \emph{translation} component of~$\tau$.
          \item $\nu_2$ is chosen as either the current position of the leader~$\Ltwo$ or, if $\Ltwo$ already lies on~$\nu_1$, the point $(0,1)$ in the local coordinate system of~$\Ltwo$.  
                By construction $\nu_2$ must belong to the image under~$\tau$ of (one of) the class $[v_{\iota_1}]_\equiv$ that lies on $SEC(\Pi_i)$.
                Hence $\dist{\nu_1,\nu_2}$ uniquely fixes the non-zero \emph{scale} of~$\tau$.
          \item The leader~$\Lthree$ selects unambiguously a multivertex of $\Pi_i$ and moves to it.
                If $\Pi_i\notin\NLine$ then such a multivertex is chosen so that it is not aligned with $\Lone$ and $\Ltwo$.
                Thus, the oriented segment $\overline{\nu_1\nu_3}$ therefore fixes the \emph{rotation} (and hence the chirality) of~$\tau$.
                If $\Pi_i\in\NLine$, no rotation is needed.
        \end{itemize}

              The three conditions above determine a unique similarity
              transformation $\tau$ and therefore a unique pattern
              $\Pi=\tau(\Pi_i)$.  No other choice of $\Pi$
              is admissible without violating at least one of the
              selection rules (radius, multivertex class, or oriented
              order), so $\triangle\nu_1\nu_2\nu_3$ identifies exactly
              one representative of $\Pats{\Pi_i}$.
    \end{itemize}
\end{proof}

\subsection{Correctness of \Phase{2}}
Let $\Pi\in \Pats{\Pi_i}$ be the actual pattern to be formed during \Phase{2}.

\begin{lemma}\label{lem:invariantphase2}
    During \Phase{2}, $\Pi$ is invariant.
\end{lemma}
\begin{proof}
    By \Cref{lemma:pi_unambiguous}, $\Pi$ is unequivocally defined by the chiral angle, which is in turn defined by the positions of the leaders $\Lone$, $\Ltwo$, $\Lthree$.
    During \Phase{2}, the leaders do not move, except for $\Lone$ when it moves from $\centersec{C}$ to reach its target vertex, if $\Pi_i\in \NPoints$.
    Thus, if $\Pi_i\notin \NPoints$, $\triangle \Lone\Ltwo\Lthree$ defines the chiral angle.
    Otherwise, $\triangle\centersec{C}\Ltwo\Lthree$ defines the chiral angle.
    This results in $\Pi$ remaining invariant.
\end{proof}

\begin{lemma}[\footnote{For completeness, we prove the following properties on boolean words, even if probably known.}]\label{lemma:strings_first_last_different}
    Let $w=w_1\dots w_{m}$ be a minimal boolean balanced string with $m\geq 1$.
    Then, $w_1 = 1 - w_m$.
\end{lemma}
\begin{proof}
    By contradiction, assume $w_1=w_m$.
    Assume $w_1=w_m=0$ (the proof for $w_1=w_m=1$ is similar).
    We know that $w$ is a minimal balanced string; thus, for any $x$ non-trivial prefix of $w$, we have that $|x|_0 > |x|_1$ (otherwise, $w$ could be factorized into balanced strings).
    However, since $w_m=0$, then $w_1\dots w_{m-1}$ is a non-trivial prefix of $w$ containing more 1s than 0s.
    Contradiction achieved.    
\end{proof}

\begin{lemma}\label{lemma:w_factors_dyck_words}
    A balanced boolean string $w\neq \epsilon$ can be factorized in an unambiguous way into minimal-length non-null factors so that each one is either a Dyck word or the reverse of a Dyck word.
\end{lemma}
\begin{proof}
    By definition, $w\in\{0,1\}^+$ such that $|w|_0=|w|_1$.
    The factorization algorithm starts searching for the minimal-length non-null balanced prefix of $w$.
    Indeed, such a prefix, say $x$, exists, and in the worst case it corresponds to $w$ itself.
    If $w=xy$ with $y$ not empty, it is evident that $y$ is balanced.
    Thus, we can proceed by searching for the next prefix of $y$, until only the empty word remains.

    Let $x$ be a factor of $w$, resulting from this process.
    Let us consider the case when $x$ starts with 0 (i.e., the opening bracket). 
    We have to prove that $x$ is a Dyck word.
    We already know that it is balanced.
    We here prove that, for each non-trivial prefix of $x$, the number of 0s is $>$ than the number of 1s, which is a sufficient condition to state that $x$ is a Dyck word.
    By contradiction, let us assume that instead there exists a non-trivial prefix of $x$ for which the number of 0s is $\leq$ than the number of 1s.
    Thus, there is a non-trivial prefix of $x$ which is balanced.
    This contradicts the hypothesis that $x$ is a minimal-length non-null balanced prefix of $w$.    

    If $x$ starts with 1 (i.e., the closing bracket), then we know that $x^R$ starts with 0 by \Cref{lemma:strings_first_last_different}.
    Thus, applying the same argument, we state that $x^R$ is a Dyck word.
\end{proof}

\begin{lemma}\label{lemma:dyck_shuffle}
    Let $x=x_1\dots x_m$ be a Dyck word, and let $x_a=0$ be matched with $x_b=1$, with $1\leq a<b\leq m$.
    Let $c\in [a+1,b-1]$.
    Let $x'$ be the string obtained by moving $x_b$ from the $b$-th position to the $c$-th position, and shifting all the other symbols, i.e.,
        $$x'_i = \begin{cases}
                    x_i& \text{if } i<c \land i >b\\
                    x_b & \text{if } i= c\\
                    x_{i-1} & \text{if } c<i \leq b.
                \end{cases}$$

    Then, $x'$ is a Dyck word.
    Moreover, the matched $0$ of $x'_c=1$ is contained in the factor $x'_a\cdots x'_{c-1}$.
\end{lemma}
\begin{proof}
    Indeed, $x'$ is balanced.
    If $x'=x$, then trivially the property is true.
    Otherwise, we have to prove that $\delta_{x'}(i) \geq 0$ for any $i$.
    To lighten the notation, we use $\delta(\cdot)$ in place of $\delta_x(\cdot)$, and $\delta'(\cdot)$ in place of $\delta_{x'}(\cdot)$.
    We have that:
    \begin{equation}\label{eq:delta}
        \delta'(i) = \begin{cases}
        \delta(i) & \text{if } i<c \land i >b\\
        \delta(i-1) -1 & \text{if } c\leq i \leq b
    \end{cases}
    \end{equation}
    Note that, since $x_a$ and $x_b$ are matching brackets in $x$, then $\delta(a)=\delta(b)+1 > 0$, and it holds that $\delta(i) >\delta(a)-1$ for any $a<i<b$ (otherwise, $x_a$ and $x_b$ would not match).
    This proves that $\delta'\geq 0$ in all the cases.

    We have now to prove that $x'_c$ (which is the moved 1) is matched with a 0 in $x'_a\cdots x'_{c-1}$.
    By contradiction, assume that $x'_c$ is matched with a 0 in the prefix $x'_1\cdots x'_{a-1}$.
    Thus, $x'_a=0$ is matched with a 1 in $x'_{a+1}\cdots x'_{c-1}$.
    However, since $x'_1\cdots x'_{c-1} = x_1\cdots x_{c-1}$, such a matching would have existed also in $x$.
    Contradiction achieved.
\end{proof}

\begin{lemma}\label{lemma:dyck_matching_still_forward}
    Let $r$ be a robot lying at a point $a\in D$ during \Phase{2}.
    Assume $r$ is activated and has to reach its target projection $v\in D$.
    Assume $r$ is stopped during its trajectory at position $a'$ in the open segment $\overline{va}$ (i.e., $a'\in\overline{va}\setminus\{v,a\}$).
    Then, the (possibly new) target projection of $r$ belongs to the closed segment $\overline{va'}$.
\end{lemma}
\begin{proof}
    If $a'$ corresponds to a free vertex, then $a'$ becomes the new target projection of $r$ by construction (thus $r$ must stay still).
    Otherwise, we have to prove that, according to the new arrangement string, robot $r$ is assigned to a vertex (thus 0) in the substring corresponding to the segment $\overline{va'}$.
    Let $w=w_1\cdots w_{m}$ be the arrangement string before the movement of $r$, which thus matches $a$ with $v$, and let $w=y^{(1)}\cdots y^{(h)}$ be the factorization in minimal balanced strings, with $h\geq 1$.
    Let $y^{(i)}$ be the minimal balanced factor that contains $a$ and $v$.
    W.l.o.g., assume that $y^{(i)}$ is a Dyck word (otherwise, consider its reverse, by \Cref{lemma:w_factors_dyck_words}).
    Let $y'$ be the new balanced factor, obtained after the movement of $r$ to $a'$ (note that the other factors are not affected by this movement).
    By \Cref{lemma:dyck_shuffle}, we know that $y'$ is a Dyck word, and that the 1 corresponding to $a'$ is matched to a 0 in the factor corresponding to $\overline{va'}$.    
\end{proof}

\begin{lemma}\label{lemma:non_rigid_Lone}
No ambiguity arises if $\Lone$ is stopped by the adversary while traveling during \Phase{1} (towards the center of the configuration) or during \Phase{2} (towards its target vertex).
\end{lemma}
\begin{proof}
    Suppose that $\Lone$ is stopped by the adversary before reaching its target destination.
    Let $C_{|\Reals^2}\in\NPoints$ be this transitory configuration (i.e., where the swarm does not form $\Pi_i$ and $\Lone$ does not lie on $\centersec{C}$).
    Note that a configuration with these properties can occur both during \Phase{1} when the leaders are moving to form the chiral angle, and during \Phase{2} when $\Lone$ is moving to reach the last vertex of the pattern, say $w$.

    Note also that, since the $w$ for $\Lone$ is chosen from the closest vertices of $\Pi\in\Pats{\Pi_i}$ to $\centersec{C}$, then $\Lone$ can form a multiplicity during its trajectory $\overline{w\centersec{C}}$ only when it stops on $w$ and $w$ is a multivertex with cardinality $>1$.
    Thus, when $\Lone$ is stopped before reaching $w$ (end of \Phase{2}) or before reaching $\centersec{C}$ (beginning of \Phase{1}), it does not form multiplicities.

    We now prove that the two situations are unambiguous.
    If any other robot (but $\Lone$) is activated in $C$, it will see that $\Lone$ does not lie on $\centersec{C}$, thus, it will stay still by construction.
    As soon as $\Lone$ is reactivated, it computes the counter $i$, and it must understand if it was moving towards $\centersec{C}$ or towards the last vertex of the similar pattern to $\Pi_i$.
    For this purpose, it computes the chiral angle $\triangle \nu_1\nu_2\nu_3$ where $\nu_1=\centersec{C}$, and thus the actual pattern $\Pi$ to be formed.
    If the other robots of the swarm cover all the vertices of $\Pi$, and only one vertex $w$ is uncovered, and the open line segment $\overline{w\Lone}$ is robot-free, then $\Lone$ is aware that it must head to $w$ (\Phase{2}).
    Otherwise, it heads to $\centersec{C}$ in order to form the chiral angle for the new pattern (\Phase{1}).
\end{proof}

\subsection{Runtime complexity}

Runtime complexity is measured in epochs and, since we consider non-rigid movements, it depends on the non-rigidity parameter $\delta$.
We indicate with $\wtime(\cdot)$ the number of epochs taken to accomplish a given phase of the algorithm.
We denote the radius of a circle with $\radius{\cdot}$.

\begin{lemma}\label{lemma:time_phase_0}
For every pattern $\Pi_i\in \choreo$, $\wtime(\Phase{0}) \leq 3$.
\end{lemma}
\begin{proof}
    During \Phase{0}, no robot moves, and only three $\off$ robots must elect as leaders and change their colors properly.
    Thus, this phase lasts at most three epochs, assuming, for simplicity, that leaders elect themselves in sequence (i.e., $\Ltwo$, $\Lone$, and $\Lthree$).
\end{proof}

\begin{lemma}\label{lemma:time_phase_1}
    Let $C$ be the configuration from which \Phase{1} starts for a pattern $\Pi_i\in \choreo$.
    If $C_{|\Reals^2}\in\Point$, let $C'$ be the configuration obtained after the movement of $\Ltwo$.
    Let $\Omega$ be $SEC(C)$ if $C_{|\Reals^2}\in\NPoints$, otherwise let it be the circle centered in $pos(\Lone)$ and whose radius is $\dist{\Lone,\Ltwo}$ if $C,C'\in (\TwoPoints\cup \ThreePoints)$.
    Then,
    \begin{equation}
        \wtime(\Phase{1})\leq 
            \begin{cases}
                \frac{3\radius{\Omega}}{\delta} & \text{if } \Pi_i \in \Point \\
                \frac{2\radius{\Omega}}{\delta} & \text{if } \Pi_i \notin \Point.
            \end{cases} 
    \end{equation}
\end{lemma}
\begin{proof}
We remind that during \Phase{1} the leaders $\Lone$, $\Ltwo$, and $\Lthree$ move to form the chiral angle $\triangle\nu_1\nu_2\nu_3$ for defining the pattern $\Pi\in \Pats{\Pi_i}$ to be formed.
By construction, $\nu_1$ always corresponds to the center of $\Omega$.
Let us consider the four cases:
\begin{itemize}
    \item If $\Pi_i\in \Point$, then $\nu_1=\nu_2=\nu_3=\centersec{\Omega}$. 
    Note that $C_{|\Reals^2}\notin\Point$, otherwise the pattern is already formed.
    Thus, the maximum traveled distance occurs when all the leaders must reach $\centersec{\Omega}$, by traveling each one a distance $\radius{\Omega}$: in total, this case takes $\frac{3\radius{\Omega}}{\delta}$ epochs.
    \item If $\Pi_i\notin\Point$, the worst case is when both $\Lone$ and $\Lthree$ must reach $\nu_1$ $\nu_3$, each one traveling the maximum distance $\radius{\Omega}$.
    Thus, the worst case takes $\frac{2\radius{\Omega}}{\delta}$ epochs.
\end{itemize}
\end{proof}

\begin{lemma}\label{lemma:time_phase_2}
    Let $C$ be the configuration at the beginning of \Phase{2}, where $\triangle\nu_1\nu_2\nu_3$ is the chiral angle for $\Pi_i$.
    Let $\Omega$ be the circle having center in $\nu_1$ and whose radius is $\max_{r\in\swarm}\{\dist{\nu_1,r}\}$.
    Then,
    \begin{equation*}
        \wtime(\Phase{2})\leq 
            \begin{cases}
                \frac{\radius{\Omega}}{\delta} & \text{if } \Pi_i \in \Point \\
                \frac{2\radius{\Omega}}{\delta}& \text{if } \Pi_i \in (\TwoPoints\cup \ThreePoints) \\
                \frac{5\radius{\Omega}}{\delta} & \text{if } \Pi_i \in \NPoints
            \end{cases} 
    \end{equation*}
\end{lemma}
\begin{proof}
    Note that $\Omega$ corresponds to the supporting circle of $\Pi_i$ if and only if $\Pi_i\notin\Point$.
    If $\Pi_i\in\Point$, all the robots must gather on $\nu_1$, thus in the worst case each robot travels a distance $\radius{\Omega}$.
    If $\Pi_i \in (\TwoPoints\cup \ThreePoints)$, the maximum distance traveled by a robot corresponds to the diameter of $\Omega$.
    Otherwise ($\Pi_i \in \NPoints$), the algorithm makes all non-leader robots arrange on the $n-3$ vertices of the pattern in three sub-phases, and eventually makes $\Lone$ reach the last missing vertex.
    In the three sub-phases, the non-leaders reach the $D$ (maximum traveled distance $\radius{\Omega}$), then they arrange on $D$ covering the vertices' projections (maximum traveled distance $2\radius{\Omega}$), and lastly they travel perpendicularly to reach their target vertices (maximum traveled distance $\radius{\Omega}$).
    Eventually, $\Lone$ reaches the last vertex (maximum distance $\radius{\Omega}$).
    Thus, if $\Pi_i \in \NPoints$, the phase takes $\frac{5\radius{\Omega}}{\delta}$ epochs.
\end{proof}

\begin{lemma}\label{lemma:time_phase_3}
    $\wtime(\Phase{3}) \leq 3$.
\end{lemma}
\begin{proof}
    During \Phase{3}, no robot moves.
    Let $\choreo = (\Pi_0,\dots, \Pi_{q-1})^x$ be the choreography the swarm is performing.
    Let $i\in[0,q-1]$ be the index encoded by the swarm through the Gray code.
    If $x=1$ or if $x=\infty$ and $i<q-1$, then this phase takes only one epoch.
    In fact, at most one robot has to change its color to update the counter value from $i$ to $i+1$.

    Otherwise, it takes one epoch for $\Lthree$ to set to $\Ltwo$, one for all $n-3$ non-leaders to set to $\off$, and the last one for the resetting of $\Lthree$.
\end{proof}

\subsection{Spatial homogeneity}
Our strategy for constructing the supporting circles is to guarantee a sort of spatial homogeneity in the performance of $\choreo = (\Pi_0,\dots, \Pi_{q-1})^x$. 
Notably,

\begin{lemma}
    Let $\Omega_0,\Omega_1,\Omega_2,\dots$ be the sequence of supporting circles formed along the performance of $\choreo$, so that $\Omega_j$ is the supporting circle for $\Pi_{j\pmod q}$.
    Then,
    \begin{itemize}
        \item $\Omega_0,\Omega_1,\Omega_2,\dots$ are all concentric;
        \item given any subsequence $\Omega_{a},\Omega_{a+1},\dots,\Omega_{b}$ such that $\Pi_{i\pmod q}\notin\Point$ for any $i\in[a,b]$, then $\Omega_{a},\Omega_{a+1},\dots,\Omega_{b}$ are equal.
    \end{itemize}
\end{lemma}
\begin{proof}
    Let us prove both the claims:
    \begin{itemize}
        \item Let $\omega$ be the center of $\Omega_0$. 
        If $\Pi_0\in \NPoints$, then the pattern $\Pi\in\Pats{\Pi_0}$ will be formed so that $SEC(\Pi) = \Omega_0$, and once formed, $\Lone$ will come back to $\omega$ to form the (the vertex $\nu_1$ of the) chiral angle for $\Pi_1$. 
        Instead, if $\Pi_0\notin \NPoints$, $\Lone$ will never move for the formation of the pattern $\Pi_0$, and $pos(\Lone)$ constitutes a vertex of $\Pi_0$.
        Thus, in both cases, the next supporting circle $\Omega_1$ will be centered in $\omega$ as well.
        Recursively, the claim holds for the entire subsequence.
        \item By construction, the supporting circle of the pattern $\Pi_{a\pmod q}\notin\Point$ is non-degenerate, and it is given by $pos(\Lone)$ (center) and $pos(\Ltwo)$ (radius) after the setting of the corresponding chiral angle. 
        After the formation of $\Pi_{a\pmod q}$, the leaders must form the chiral angle (and thus the supporting circle) for $\Pi_{a+1\pmod q}$.
        By the first claim, we know that $\Omega_{a+1}$ has the same center as $\Omega_a$. 
        Since $\Ltwo$ only moves when the swarm must form a $\Point$, then it results that the two circles have the same radius.
        Recursively, the claim holds for the entire subsequence.
    \end{itemize}
\end{proof}

\subsection{Number of colors}
The algorithm here proposed uses $k\geq 4$ colors: three are used by the leaders (i.e., $\Lone,\Ltwo,\Lthree$) while the remaining $k'=k-3$ colors are used by the $n-3$ non-leader robots to implement the counter.
Thus, exploiting the technique of Gray codes as explained in \Cref{sec:graycodes}, the number of possible composition vectors is $\binom{n-3+k'-1}{k'-1} = \binom{n+k-7}{k-4}$.
This allows to perform any choreography of lenght/period $q \leq \binom{n+k-7}{k-4}$.

    \section{Extra material}\label{appendix:extra}
\subsection{Pseudo-code}\label{axsec:pseudocode}
\Cref{algo:pseudocode} represents the pseudo-code of the whole algorithm presented in this paper.

\begin{algorithm}[h]
	\caption{Pseudocode.}
        \label{algo:pseudocode}
        {\small	
        \SetKwInput{Input}{Input}
        \SetKwInput{Output}{Output}
        \SetKwInput{Result}{Final Result}
        
        \SetKwBlock{PHzero}{\Phase{0}}{end}
        \SetKwBlock{PHone}{\Phase{1}}{end}
        \SetKwBlock{PHtwo}{\Phase{2}}{end}
        \SetKwBlock{PHthree}{\Phase{3}}{end}

        \Input{A swarm $\swarm=\{r_1,\dots,r_n\}$ which has to perform $\choreo = (\Pi_0, \dots, \Pi_{q-1})^{x\in \{1, \infty\}}$}
        \PHzero{
            \Input{$C$ initial configuration with $n$ $\off$ robots on $\Reals^2$}
            Color setting of the leaders: $\Ltwo$, $\Lone$, and $\Lthree$\;
        }
        \While{true}{
            $i\gets $ value of counter\;
            $\Pi_i \gets $ pattern to be formed\;
            \PHone{
                \Input{$C$ current configuration}
                \If{$C_{|\Reals^2}\notin \Pats{\Pi_i}$}{
                    $\Lone,\Ltwo,\Lthree$ set the chiral angle $\triangle \nu_1\nu_2\nu_3$ for $\Pi_i$\;
                }
            }
            \PHtwo{
               \Input{$C$ current configuration}
               
                \If{$C_{|\Reals^2}\neq \Pi$ and $\Lone,\Ltwo,\Lthree$ form $\triangle \nu_1\nu_2\nu_3$ for $\Pi_i$}{
                    \If{$\Pi\notin\NPoints$}{
                        $r$ reaches the closest unsaturated multivertex of $\Pi$;
                    }
                    \Else{
                        $D \gets$ main diameter\;
                        $\swarm_{NL} = \swarm \setminus\{\Lone,\Ltwo,\Lthree\}$ non-leader robots\;
                        $\swarm_{NL}$ migrates on $D$ perpendicularly\;
                         $\swarm_{NL}$ arranges on the vertex projections on $D$\;
                         $\swarm_{NL}$ reaches the target vertices moving perpendicularly to $D$\;
                        $\Lone$ reaches the last missing vertex\;
                    }
                }
            }
            \PHthree{
                $\Pi \gets $ pattern in $\Pats{\Pi_i}$ defined by $\triangle \nu_1\nu_2\nu_3$\;
                \If{$C_{|\Reals^2} =\Pi$}{
                    \If{$i=q-1$}{
                        \If{$x = \infty$}{
                            $\Lthree$ sets its color to $\Ltwo$\;
                            $\swarm_{NL}$ sets the counter to 0\;
                            resetting of $\Lthree$\;
                        }
                    }
                    \Else{
                        $\swarm_{NL}$ sets the counter to $i+1$\;
                    }
                    }
                } 

        } 
        }
\end{algorithm}

\subsection{Figures and tables}\label{axsec:figures}
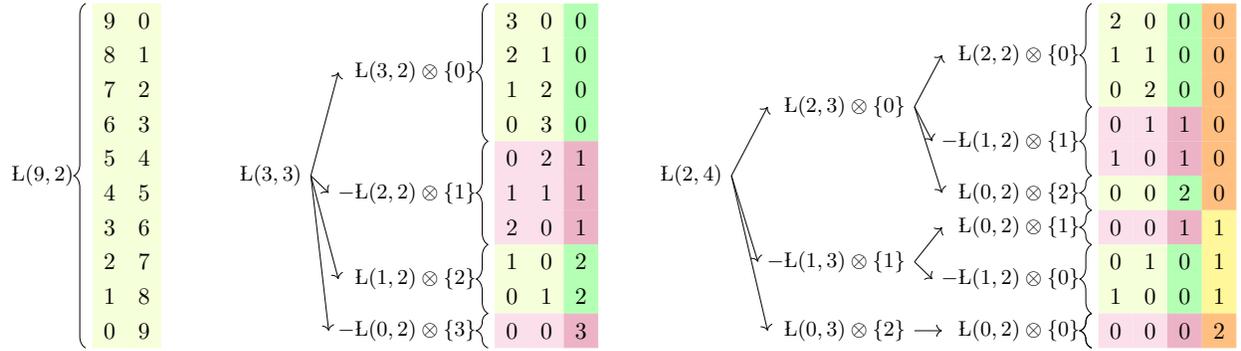
\begin{figure}[h]
\centering
\resizebox{\textwidth}{!}{%
\begin{tikzpicture}

\tikzset{
    box/.style={
        minimum width=0.5cm,
        minimum height=0.5cm,
        inner sep=1pt,
        font=\small
    },
    base/.style={box, fill=lime!15, draw=lime!15},
    baserev/.style={box, fill=magenta!15, draw=magenta!15},
    firstRec/.style={box, fill=green!30, draw=green!30},
    firstRecs/.style={box, fill=purple!30, draw=purple!30},
    secondRec/.style={box, fill=orange!50, draw=orange!50},
    secondRecs/.style={box, fill=yellow!50, draw=yellow!50},
    treeNode/.style = {node distance=2cm, sibling distance=2cm}
}

\node[base, below] (L92-1) at (0,10) {9};
\node[base, right=0 of L92-1] (L92-2) {0};

\node[base, below=0 of L92-1] (L92-3) {8};
\node[base, right=0 of L92-3] (L92-4) {1};

\node[base, below=0 of L92-3] (L92-5) {7};
\node[base, right=0 of L92-5] (L92-6) {2};

\node[base, below=0 of L92-5] (L92-7) {6};
\node[base, right=0 of L92-7] (L92-8) {3};

\node[base, below=0 of L92-7] (L92-9) {5};
\node[base, right=0 of L92-9] (L92-10) {4};

\node[base, below=0 of L92-9] (L92-11) {4};
\node[base, right=0 of L92-11] (L92-12) {5};
\node[base, below=0 of L92-11] (L92-13) {3};
\node[base, right=0 of L92-13] (L92-14) {6};
\node[base, below=0 of L92-13] (L92-15) {2};
\node[base, right=0 of L92-15] (L92-16) {7};
\node[base, below=0 of L92-15] (L92-17) {1};
\node[base, right=0 of L92-17] (L92-18) {8};
\node[base, below=0 of L92-17] (L92-19) {0};
\node[base, right=0 of L92-19] (L92-20) {9};

\node[left=0.4cm] at ($(L92-1)!0.5!(L92-19)$) {\footnotesize $\L(9,2)$};

\draw[decorate, decoration={brace, mirror, amplitude=5pt}]
  ([xshift=-0.1cm]L92-1.north west) -- ([xshift=-0.1cm]L92-19.south west);

\node[base, below] (L33-1) at (6,10) {3};
\node[base, right=0 of L33-1] (L33-2) {0};
\node[firstRec, right=0 of L33-2] (L33-3) {0};

\node[base, below=0 of L33-1] (L33-4) {2};
\node[base, right=0 of L33-4] (L33-5) {1};
\node[firstRec, right=0 of L33-5] (L33-6) {0};

\node[base, below=0 of L33-4] (L33-7) {1};
\node[base, right=0 of L33-7] (L33-8) {2};
\node[firstRec, right=0 of L33-8] (L33-9) {0};

\node[base, below=0 of L33-7] (L33-10) {0};
\node[base, right=0 of L33-10] (L33-11) {3};
\node[firstRec, right=0 of L33-11] (L33-12) {0};

\node[left=0.4cm] at ($(L33-1)!0.5!(L33-10)$) (L32x0) {\footnotesize $~\L(3,2) \otimes \{0\}$};
\draw[decorate, decoration={brace, mirror, amplitude=5pt}]
  ([xshift=-0.1cm]L33-1.north west) -- ([xshift=-0.1cm]L33-10.south west);

\node[baserev, below=0 of L33-10] (L33-13) {0};
\node[baserev, right=0 of L33-13] (L33-14) {2};
\node[firstRecs, right=0 of L33-14] (L33-15) {1};

\node[baserev, below=0 of L33-13] (L33-16) {1};
\node[baserev, right=0 of L33-16] (L33-17) {1};
\node[firstRecs, right=0 of L33-17] (L33-18) {1};

\node[baserev, below=0 of L33-16] (L33-19) {2};
\node[baserev, right=0 of L33-19] (L33-20) {0};
\node[firstRecs, right=0 of L33-20] (L33-21) {1};

\node[left=0.4cm] at ($(L33-13)!0.5!(L33-19)$) (L22x1) {\footnotesize $-\L(2,2) \otimes \{1\}$};
\draw[decorate, decoration={brace, mirror, amplitude=5pt}]
  ([xshift=-0.1cm]L33-13.north west) -- ([xshift=-0.1cm]L33-19.south west);

\node[base, below=0 of L33-19] (L33-22) {1};
\node[base, right=0 of L33-22] (L33-23) {0};
\node[firstRec, right=0 of L33-23] (L33-24) {2};

\node[base, below=0 of L33-22] (L33-25) {0};
\node[base, right=0 of L33-25] (L33-26) {1};
\node[firstRec, right=0 of L33-26] (L33-27) {2};

\node[baserev, below=0 of L33-25] (L33-28) {0};
\node[baserev, right=0 of L33-28] (L33-29) {0};
\node[firstRecs, right=0 of L33-29] (L33-30) {3};

\node[left=0.4cm] at ($(L33-22)!0.5!(L33-25)$) (L12x2) {\footnotesize $~\L(1,2) \otimes \{2\}$};
\node[left=0.4cm] at (L33-28) (L02x3) {\footnotesize $-\L(0,2) \otimes \{3\}$};
\draw[decorate, decoration={brace, mirror, amplitude=5pt}]
  ([xshift=-0.1cm]L33-22.north west) -- ([xshift=-0.1cm]L33-25.south west);
\draw[decorate, decoration={brace, mirror, amplitude=5pt}]
  ([xshift=-0.1cm]L33-28.north west) -- ([xshift=-0.1cm]L33-28.south west);

\node[left=3cm] at ($(L33-1)!0.5!(L33-28)$) (L33) {\footnotesize $\L(3,3)$};

\draw[->] (L33.east) -- (L32x0.west);
\draw[->] (L33.east) -- (L22x1.west);
\draw[->] (L33.east) -- (L12x2.west);
\draw[->] (L33.east) -- (L02x3.west);


\node[base, below] (L24-1) at (15,10) {2};
\node[base, right=0 of L24-1] (L24-2) {0};
\node[firstRec, right=0 of L24-2] (L24-3) {0};
\node[secondRec, right=0 of L24-3] (L24-4) {0};

\node[base, below=0 of L24-1] (L24-5) {1};
\node[base, right=0 of L24-5] (L24-6) {1};
\node[firstRec, right=0 of L24-6] (L24-7) {0};
\node[secondRec, right=0 of L24-7] (L24-8) {0};

\node[base, below=0 of L24-5] (L24-9) {0};
\node[base, right=0 of L24-9] (L24-10) {2};
\node[firstRec, right=0 of L24-10] (L24-11) {0};
\node[secondRec, right=0 of L24-11] (L24-12) {0};

\node[left=0.4cm] at ($(L24-1)!0.5!(L24-9)$) (L22x0_24) {\footnotesize $~\L(2,2) \otimes \{0\}$};
\draw[decorate, decoration={brace, mirror, amplitude=5pt}]
  ([xshift=-0.1cm]L24-1.north west) -- ([xshift=-0.1cm]L24-9.south west);

\node[baserev, below=0 of L24-9] (L24-13) {0};
\node[baserev, right=0 of L24-13] (L24-14) {1};
\node[firstRecs, right=0 of L24-14] (L24-15) {1};
\node[secondRec, right=0 of L24-15] (L24-16) {0};

\node[baserev, below=0 of L24-13] (L24-17) {1};
\node[baserev, right=0 of L24-17] (L24-18) {0};
\node[firstRecs, right=0 of L24-18] (L24-19) {1};
\node[secondRec, right=0 of L24-19] (L24-20) {0};

\node[left=0.4cm] at ($(L24-13)!0.5!(L24-17)$) (L12x1_24) {\footnotesize $-\L(1,2) \otimes \{1\}$};
\draw[decorate, decoration={brace, mirror, amplitude=5pt}]
  ([xshift=-0.1cm]L24-13.north west) -- ([xshift=-0.1cm]L24-17.south west);

\node[base, below=0 of L24-17] (L24-21) {0};
\node[base, right=0 of L24-21] (L24-22) {0};
\node[firstRec, right=0 of L24-22] (L24-23) {2};
\node[secondRec, right=0 of L24-23] (L24-24) {0};

\node[left=0.4cm] at ($(L24-21)!0.5!(L24-21)$) (L02x2_24) {\footnotesize $~\L(0,2) \otimes \{2\}$};
\draw[decorate, decoration={brace, mirror, amplitude=5pt}]
  ([xshift=-0.1cm]L24-21.north west) -- ([xshift=-0.1cm]L24-21.south west);

\node[baserev, below=0 of L24-21] (L24-25) {0};
\node[baserev, right=0 of L24-25] (L24-26) {0};
\node[firstRecs, right=0 of L24-26] (L24-27) {1};
\node[secondRecs, right=0 of L24-27] (L24-28) {1};

\node[base, below=0 of L24-25] (L24-29) {0};
\node[base, right=0 of L24-29] (L24-30) {1};
\node[firstRec, right=0 of L24-30] (L24-31) {0};
\node[secondRecs, right=0 of L24-31] (L24-32) {1};

\node[base, below=0 of L24-29] (L24-33) {1};
\node[base, right=0 of L24-33] (L24-34) {0};
\node[firstRec, right=0 of L24-34] (L24-35) {0};
\node[secondRecs, right=0 of L24-35] (L24-36) {1};

\node[baserev, below=0 of L24-33] (L24-37) {0};
\node[baserev, right=0 of L24-37] (L24-38) {0};
\node[firstRecs, right=0 of L24-38] (L24-39) {0};
\node[secondRec, right=0 of L24-39] (L24-40) {2};

\node[left=3cm] at ($(L24-1)!0.5!(L24-21)$) (L23x0_24) {\footnotesize $~\L(2,3)\otimes\{0\}$};

\node[left=3cm] at ($(L24-25)!0.5!(L24-33)$) (L13x1_24) {\footnotesize $-\L(1,3) \otimes \{1\}$};
\node[left=3cm] at ($(L24-37)!0.5!(L24-37)$) (L03x2_24) {\footnotesize $~\L(0,3) \otimes \{2\}$};

\node[left=0.4cm] at ($(L24-25)!0.5!(L24-25)$) (L02x1_24) {\footnotesize $~\L(0,2) \otimes \{1\}$};
\node[left=0.4cm] at ($(L24-29)!0.5!(L24-33)$) (L12x0_24) {\footnotesize $-\L(1,2) \otimes \{0\}$};
\node[left=0.4cm] at ($(L24-37)!0.5!(L24-37)$) (L02x0_24) {\footnotesize $~\L(0,2) \otimes \{0\}$};

\node[left=6.5cm] at ($(L24-1)!0.5!(L24-40)$) (L24) {\footnotesize $\L(2,4)$};
\draw[->] (L23x0_24.east) -- (L22x0_24.west);
\draw[->] (L23x0_24.east) -- (L12x1_24.west);
\draw[->] (L23x0_24.east) -- (L02x2_24.west);
\draw[->] (L13x1_24.east) -- (L02x1_24.west);
\draw[->] (L13x1_24.east) -- (L12x0_24.west);
\draw[->] (L03x2_24.east) -- (L02x0_24.west);

\draw[->] (L24.east) -- (L23x0_24.west);
\draw[->] (L24.east) -- (L13x1_24.west);
\draw[->] (L24.east) -- (L03x2_24.west);

\draw[decorate, decoration={brace, mirror, amplitude=5pt}]
  ([xshift=-0.1cm]L24-37.north west) -- ([xshift=-0.1cm]L24-37.south west);
\draw[decorate, decoration={brace, mirror, amplitude=5pt}]
  ([xshift=-0.1cm]L24-25.north west) -- ([xshift=-0.1cm]L24-25.south west);
\draw[decorate, decoration={brace, mirror, amplitude=5pt}]
  ([xshift=-0.1cm]L24-29.north west) -- ([xshift=-0.1cm]L24-33.south west);
\draw[decorate, decoration={brace, mirror, amplitude=5pt}]
  ([xshift=-0.1cm]L24-37.north west) -- ([xshift=-0.1cm]L24-37.south west);

\end{tikzpicture}
}
\caption{Gray codes for $\L(9,2)$, $\L(3,3)$, and $\L(2,4)$. They allow a group of 9, 3 or 2 robots to implement a counter from 0 to 9 by using 2, 3, or 4 colors, respectively.}
\label{fig:graycode}
\end{figure}
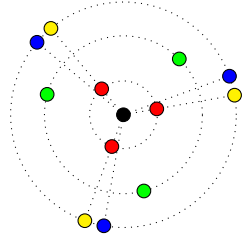
\begin{figure}[h]
    \centering
    \begin{tikzpicture}[scale=0.5, transform shape, font = {\LARGE}]
		\def\r{3cm}
            \def\a{360/3}

            \foreach \i in {1,0.7,0.3,0}{
                \draw [thin, dotted] (0,0) circle (\r*\i);
            }
            \foreach \i in {0,1,2}{
                \draw [thin, dotted] (0,0) -- (\i*\a+10:\r);
                \draw [thin, dotted] (0,0) -- (\i*\a+20:\r);

                \rnode{yellow}{above}{}{\i*\a+10:\r};
                \rnode{blue}{above}{}{\i*\a+20:\r};
                \rnode{green}{above}{}{\i*\a+45:\r*0.7};
                \rnode{red}{above}{}{\i*\a+10:\r*0.3};

            }
            \rnode{black}{above}{}{0:0};

	\end{tikzpicture}
    \caption{Ranking of the multivertices. Assuming each multivertex has cardinality 1, the classes of multivertices are ordered in this way: $[v_{yellow}] < [v_{blue}] < [v_{green}] < [v_{red}] < [v_{black}]$.}
    \label{fig:ranking_multivertices}
\end{figure}
\begin{table}[h!]
    \centering
    \resizebox{\textwidth}{!}{%
    \renewcommand{\arraystretch}{2}{
    \begin{tabular}{|c|c|c|}
        \hline
        $\Pi_i$     &  $\triangle\nu_1\nu_2\nu_3$ & $\Pi\in\Pats{\Pi_i}$ to be formed\\
        \hline
        \hline
        
        \makecell{$\in\Point$\\$=\{(v_1,n)\}$}   & $\nu_1=\nu_2=\nu_3=\begin{cases}
                                                                            \centersec{C} & C_{|\Reals^2}\in\NPoints\\
                                                                            pos(\Lone) & \text{otherwise}
                                                                            \end{cases}$ & $\Pi=\{(\nu_1,n)\}$\\
        \Xhline{2\arrayrulewidth}

        \makecell{$\in\TwoPoints$ \\
                        $=\{(v_1,\card_1),(v_2,\card_2)\}$\\ 
                         $\card_1\geq \card_2$} 

                                                    & \makecell{$\nu_1=\nu_3=\begin{cases}
                                                                            \centersec{C} & C_{|\Reals^2}\in\NPoints\\
                                                                            pos(\Lone) & \text{otherwise}
                                                                            \end{cases}$\\
                                                                $\nu_2=\begin{cases}
                                                                            (0,1) \text{ of } \Ltwo & C_{|\Reals^2}\in \Point\\
                                                                            pos(\Ltwo) &\text{otherwise}
                                                                        \end{cases}
                                                                $
                                                                }
                                                    & {$\Pi=\{(\nu_1,\card_1),(\nu_2,\card_2)\}$} \\

        \Xhline{2\arrayrulewidth}

        \makecell{$\in\ThreePoints$ \\
                        $=\{(v_1,\card_1),(v_2,\card_2),(v_3,\card_3)\}$\\ 
                        $\dist{v_1,v_2}\geq \dist{v_1,v_3}\geq \dist{v_2,v_3}$\\ with $\card_1\geq \card_2$, \\ $\card_1\geq \card_3$ if $\dist{v_1,v_3}=\dist{v_1,v_2}$}
                                                  
                                                    & \makecell{$\nu_1=\begin{cases}
                                                                            \centersec{C} & C_{|\Reals^2}\in\NPoints\\
                                                                            pos(\Lone) &\text{otherwise}
                                                                            \end{cases}$\\
                                                                $\nu_3=$ position of $v_3$ for $\Lthree$\\
                                                                $\nu_2=\begin{cases}
                                                                            (0,1) \text{ of } \Ltwo & C_{|\Reals^2}\in \Point\\
                                                                            pos(\Ltwo) & \text{otherwise}
                                                                        \end{cases}
                                                                $
                                                                }
                                                    & \makecell{$\Pi=\{(\nu_1,\card_1),(\nu_2,\card_2),(\nu_3,\card_3)\}$\\
                                                     $\dist{\nu_1,\nu_2}\geq \dist{\nu_1,\nu_3}\geq \dist{\nu_2,\nu_3}$} \\
        \Xhline{2\arrayrulewidth}
        
        \makecell{$\in\NPoints$\\
                                $=\{(v_1,\card_1),\dots,(v_m,\card_m)\}$\\
                                $[v_{\iota_1}]_\equiv <\dots < [v_{\iota_s}]_\equiv$}
                                                   
                                                    & \makecell{$\nu_1=\begin{cases}
                                                                            \centersec{C} & C_{|\Reals^2}\in\NPoints\\
                                                                            pos(\Lone) &\text{otherwise}
                                                                            \end{cases}$\\
                                                                $\nu_2=\begin{cases}
                                                                            (0,1) \text{ of } \Ltwo & C_{|\Reals^2}\in \Point\\
                                                                            pos(\Ltwo) & \text{otherwise}
                                                                        \end{cases}
                                                                $\\
                                                                $\nu_3=\begin{cases}
                                                                            \text{$v\neq\nu_1$ closest to } \nu_2 & \Pi_i\in\NLine\\
                                                                            \text{$v\neq\nu_1$ closest to } \nu_2, \text{ not aligned with $\nu_1,\nu_2$} & \text{otherwise}          
                                                                        \end{cases}
                                                                $}
                                            
                                                                &
                                                                \makecell{$\Pi=\{(\tau(v_1),\card_1),\dots,(\tau(v_m),\card_m)\}$\\
                                                                        $[\tau(v_{\iota_1})]_\equiv <\dots < [\tau(v_{\iota_s})]_\equiv$\\
                                                                        s.t. $\nu_2\in[\tau(v_{\iota_1})]_\equiv$\\
                                                                         $\nu_3\in[\tau(v)]_\equiv$}
                                                                \\
        \hline
    \end{tabular}
    }}
    \caption{Chiral angle $\triangle \nu_1\nu_2\nu_3$ and $\Pi$ in each different scenarios.}
    \label{tab:chiral_angle}
\end{table}
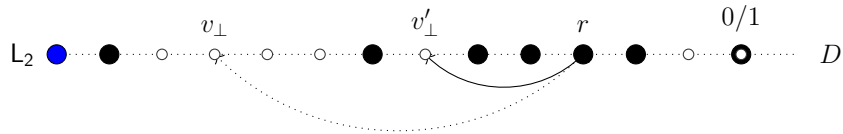
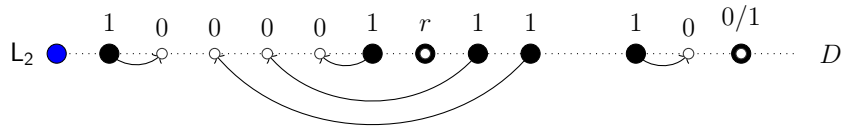
\begin{figure}[h]
    \begin{subfigure}[t]{\textwidth}
        \centering
        \begin{tikzpicture}[scale=0.7, transform shape, font = {\Large}]
        \def\l{14cm}
        \draw[dotted] (0,0) -- (\l+1,0);
        \coordinate (nu2)   at (0,0);
        \node at (\l*1.05,0)    {$D$};

        \rnode{\colorLtwo}{left}{$\Ltwo$}{nu2};

        \draw[dotted, shorten >=0.5,->]   
                    (10,0) edge      [bend left=45]        node {} (3,0)
                    ;
        \draw[thin, shorten >=0.5,->]   
                    (10,0) edge      [bend left=45]        node {} (7,0)
                    ;
        
        \foreach \x in {2,4,5,12} {
            \tnode{above}{}{\x,0};
        }
    
        \foreach \x in {1,6,8,9,11} {
            \rnode{\colorR}{above}{}{\x,0};
        }

        \rnode{\colorR}{above}{$r$}{10,0};
        \tnode{above}{$v'_\perp$}{7,0};
        \tnode{above}{$v_\perp$}{3,0};

         \rnode{\colorR}{above}{$0/1$}{13,0};
        \tnode{above}{}{13,0};

    \end{tikzpicture}
    \caption{$w=100001011110$. Trajectory computed by $r$ (dotted) vs. actual trajectory of $r$ stopped by the adversary (solid).}
    \label{subfig:dyck_word_r_stops_another_vertex}
    \end{subfigure}

    \hfill

     \begin{subfigure}[t]{\textwidth}
        \centering
        \begin{tikzpicture}[scale=0.7, transform shape, font = {\Large}]
        \def\l{14cm}
        \draw[dotted] (0,0) -- (\l+1,0);
        \coordinate (nu2)   at (0,0);
        \node at (\l*1.05,0)    {$D$};

        \rnode{\colorLtwo}{left}{$\Ltwo$}{nu2};

        \draw[shorten >=0.5,->]   
                    (1,0) edge      [bend right=45]        node {} (2,0)
                    (11,0) edge      [bend right=50]        node {} (12,0)
                    (6,0) edge      [bend left=50]        node {} (5,0)
                    (8,0) edge      [bend left=50]        node {} (4,0)
                    (9,0) edge      [bend left=50]        node {} (3,0)
                    ;
        
        \foreach \x in {2,3,4,5,12} {
            \tnode{above}{$0$}{\x,0};
        }
    
        \foreach \x in {1,6,8,9,11} {
            \rnode{\colorR}{above}{$1$}{\x,0};
        }
        \rnode{\colorR}{above}{$r$}{7,0};
        \tnode{above}{}{7,0};

         \rnode{\colorR}{above}{$0/1$}{13,0};
        \tnode{above}{}{13,0};

    \end{tikzpicture}
     \caption{New arrangement string $w=1000011110$ and new robot-projection matching after the match of $r$ with $v'_\perp$.}
    \label{subfig:dyck_word_new_matching_another_vertex}
    \end{subfigure}
    \caption{Before (a) and after (b) the non-rigid movement of $r$, which is stopped on a vertex (i.e., $v'_\perp$) different from the indented one (i.e., $v_\perp$).}
    \label{fig:dyck_word_stopped_vertex}
\end{figure}
\begin{figure}[h]
    \begin{subfigure}[t]{\textwidth}
        \centering
        \begin{tikzpicture}[scale=0.7, transform shape, font = {\Large}]
        \def\l{14cm}
        \draw[dotted] (0,0) -- (\l+1,0);
        \coordinate (nu2)   at (0,0);
        \node at (\l*1.05,0)    {$D$};

        \rnode{\colorLtwo}{left}{$\Ltwo$}{nu2};

        \draw[dotted, shorten >=0.5,->]   
                    (10,0) edge      [bend left=45]        node {} (3,0)
                    ;
        \draw[thin, shorten >=0.5,->]   
                    (10,0) edge      [bend left=45]        node {} (4.5,0)
                    ;
        
        \foreach \x in {2,4,5,7,12} {
            \tnode{above}{}{\x,0};
        }
    
        \foreach \x in {1,6,8,9,11} {
            \rnode{\colorR}{above}{}{\x,0};
        }

        \rnode{\colorR}{above}{$r$}{10,0};
        \tnode{above}{$v_\perp$}{3,0};

         \rnode{\colorR}{above}{$0/1$}{13,0};
        \tnode{above}{}{13,0};

    \end{tikzpicture}
    \caption{$w=100001011110$. Trajectory computed by $r$ (dotted) vs. actual trajectory of $r$ stopped by the adversary in a non-vertex position (solid).}
    \label{subfig:dyck_word_r_stops_no_vertex}
    \end{subfigure}

    \hfill

     \begin{subfigure}[t]{\textwidth}
        \centering
        \begin{tikzpicture}[scale=0.7, transform shape, font = {\Large}]
        \def\l{14cm}
        \draw[dotted] (0,0) -- (\l+1,0);
        \coordinate (nu2)   at (0,0);
        \node at (\l*1.05,0)    {$D$};

        \rnode{\colorLtwo}{left}{$\Ltwo$}{nu2};

        \draw[shorten >=0.5,->]   
                    (1,0) edge      [bend right=45]        node {} (2,0)
                    (11,0) edge      [bend right=50]        node {} (12,0)
                    (4.5,0) edge      [bend left=50]        node {} (4,0)
                    (8,0) edge      [bend left=50]        node {} (7,0)
                    (6,0) edge      [bend left=50]        node {} (5,0)
                    (9,0) edge      [bend left=50]        node {} (3,0)
                    ;
        
        \foreach \x in {2,3,4,5,7,12} {
            \tnode{above}{$0$}{\x,0};
        }
    
        \foreach \x in {1,4.5,6,8,9,11} {
            \rnode{\colorR}{above}{$1$}{\x,0};
        }
        \node at (4.5,-0.5)    {$r$};

        \rnode{\colorR}{above}{$0/1$}{13,0};
        \tnode{above}{}{13,0};

    \end{tikzpicture}
     \caption{New arrangement string $w=100010101110$ and new robot-projection matching.}
    \label{subfig:dyck_word_new_matching_no_vertex}
    \end{subfigure}

    \caption{Before (a) and after (b) the non-rigid movement of $r$, which is stopped on a non-vertex point.}
    \label{fig:dyck_word_stopped_no_vertex}
\end{figure}
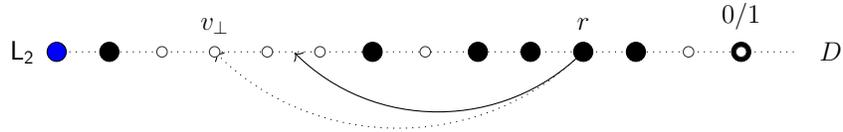
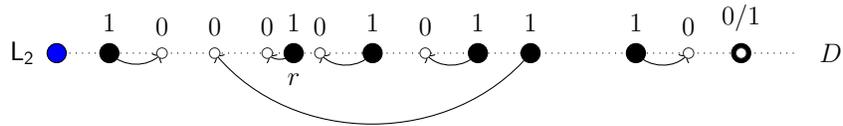

\end{document}